\documentclass[10pt,twocolumn,twoside]{IEEEtran}

\usepackage{amsmath}
\usepackage{amssymb}
\usepackage{cite}
\usepackage{graphicx}
\usepackage{ctable}

\newcommand{\BIT}{\begin{itemize}}
\newcommand{\EIT}{\end{itemize}}
\newcommand{\BEN}{\begin{enumerate}}
\newcommand{\EEN}{\end{enumerate}}

\newcommand{\M}{\mathcal{M}}
\newcommand{\B}{\mathcal{B}}

\newcommand{\X}{\mathcal{X}}

\newcommand{\Ex}{\mathbb{E}}
\providecommand{\norm}[1]{\lVert#1\rVert}
\providecommand{\ZZ}{\mathbb{Z}}
\providecommand{\bi}{\mathbf{i}}

\newcommand{\ra}{\rightarrow}

\newcommand{\non}{\nonumber}
\newcommand{\R}{\mathcal{R}}

\DeclareMathOperator{\var}{var}

\newtheorem{theorem}{Theorem}

\begin{document}
\title{Coarse Network Coding: A Simple Relay Strategy to Resolve Interference}
\author{{Peyman Razaghi}, and
    Giuseppe Caire, {\it Fellow, IEEE}\\
\thanks{ Manuscript has been submitted to the {\it IEEE Transactions on  Information Theory} on Sep.~6, 2010.  The authors are with the 
Ming Hsieh Department of Electrical Engineering, Viterbi School of Engineering,
University of Southern California,
Los Angeles, CA 90089, USA, E-mails:
{razaghi@usc.edu, caire@usc.edu}. 
Kindly please address correspondence
to Peyman Razaghi at razaghi@usc.edu.}}
\date{\today}
\maketitle
\begin{abstract}
Reminiscent of the parity function in network coding for the butterfly network, it is shown that forwarding an even/odd indicator bit for a scalar quantization of a relay  observation recovers 1 bit of information at the two destinations in a noiseless  interference channel where interference is treated as noise. Based on this  observation, a coding strategy is proposed to improve the rate of both users  at the same time using a relay node in a noisy interference channel. In this  strategy, the relay observes a linear combination of signals sent by the two  sources, and broadcasts a common message to the two destinations over a shared  out-of-band link of constant rate $R_0$ bits per channel use. The relay message  consists of the bin index of a structured binning scheme obtained from a  $2^{R_0}$-way partition of the squared lattice in the complex plane. We show that such scalar quantization-binning relay strategy asymptotically achieves the cut-set bound in an interference channel with a common out-of-band relay link of limited rate, improving  the sum rate by two bits for every bit relayed, asymptotically at high signal to  noise ratios (SNR) and when interference is treated as noise. We then use low-density parity-check (LDPC) codes along with bit-interleaved coded-modulation (BICM) as a practical coding scheme for the proposed strategy. We consider  matched and  mismatched scenarios, depending on whether the input alphabet of the interference signal is known or unknown to the decoder, respectively. 
For the matched scenario, we show the proposed strategy results in significant gains in SNR. For the mismatched scenario, we show that the proposed strategy results in rate improvements that, without the relay, cannot be achieved by merely increasing transmit powers. Finally, we use generalized mutual information analysis to characterize the theoretical performance of the mismatched scenario and validate our simulation results.
\end{abstract}

\section{Introduction}
Wireless communication is changing gears. The most promising way to meet higher and higher demands is to move towards a much higher spatial reuse, with very small cells where the distance between transmitters and receivers is dramatically reduced. On the other hand, this approach calls for a more and more unregulated user-based deployment of wireless networks, in contrast to the carefully and centrally planned conventional macro-cellular layout. In this context, femtocells (i.e., small home-based base stations) is rapidly emerging. 
Soon, people could be operating their own base stations (BS) in home or 
office, selectively controlling user access to their privately-operated BS.

Yet, interference remains as a major obstacle to widely deploy ad-hoc user-managed cellular systems, requiring further attention and new solutions. 
In a dense area, like a high-rise residential building,  inter-cell interference  is inevitable among user-controlled BSs. Because of sizable number of neighboring 
cells and small number of users in each cell,  traditional frequency reuse strategies 
are ineffective and spectrally inefficient. Hence, new strategies to handle interference are needed.

\begin{figure}
\centering
\includegraphics[width=9cm]{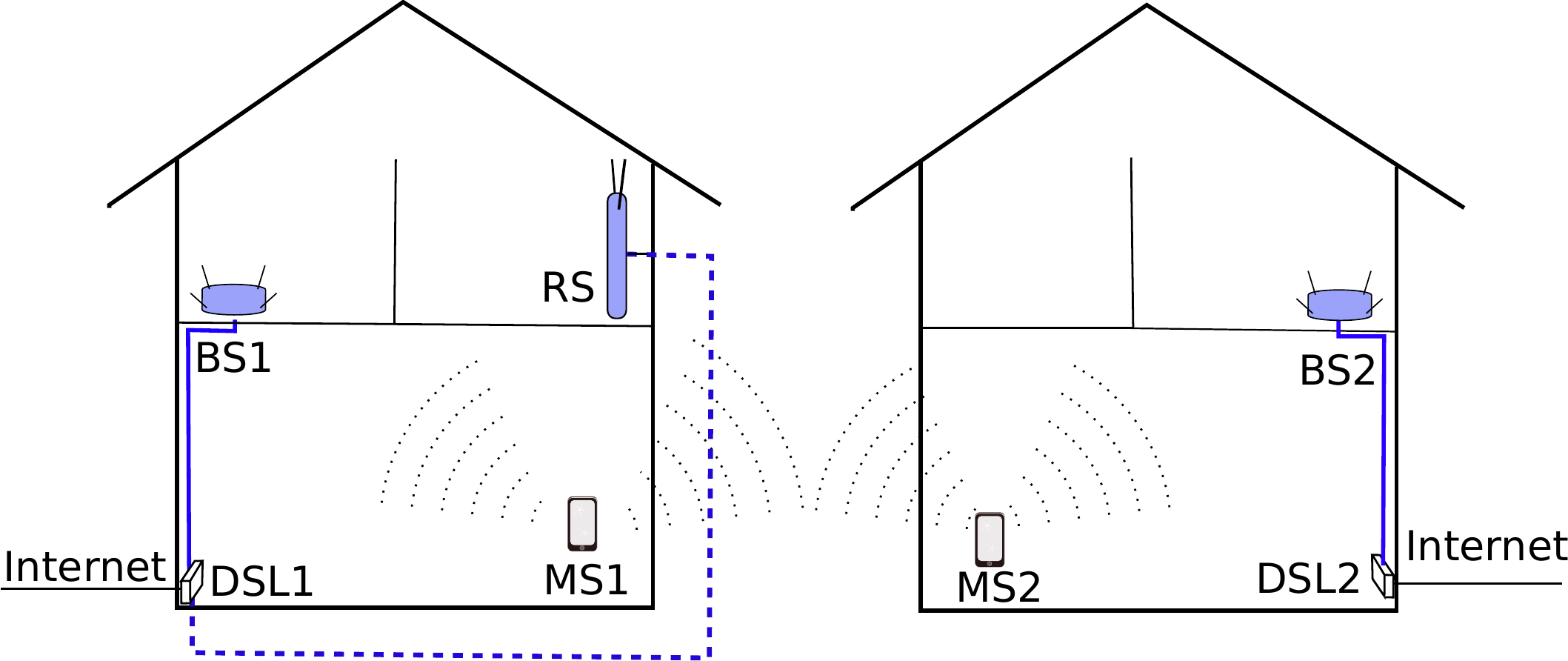}
\caption{Aside from the traditional role of data retransmission, relays can be used as ``helpers'' to resolve interference. In this figure, neighboring mobile stations (MS) in residential cellular communication networks inevitably interfere in the uplink, particularly because of restricted access policy in residential cellular systems (users cannot be distributed efficiently among BSs) and since, the stationary base stations cannot move to find a less-interfered spot. A relay station (RS) with a digital link to both base stations (BS) via Internet is used to reduce interference.  See also a similar scenario  in Fig.~\ref{fig:rcoopbs} where the relay and base stations are integrated. 
\label{fig:femtobs}}
\end{figure}
This paper introduces a simple such strategy based on relays. Relays are traditionally used as middle terminals to 
retransmit  the message.  In an interference scenario, however, we 
envision an alternative application of relays as ``helpers'' 
to resolve interference.  Fig.~\ref{fig:femtobs} shows an out-of-band 
relay placed between two neighboring interfering  cells. The relay has 
a noiseless link of a limited rate to the two destinations, and can 
broadcast a common signal to the two destinations.  The relay could 
be connected over an orthogonal wireless  connection to the two users 
in the case of downlink (e.g., WiFi) , or even wired to the two BSs  
for the uplink communication (e.g., Ethernet). 

We focus on relay strategies that involve no decoding of the user 
message. This is because the relay also experiences interference, 
rendering the decoding difficult. In our strategy, the relay operation 
involves a symbol-level quantization of the relay observation using 
a planar lattice. The relay sends a bin index for its quantized 
observation, obtained through a suitable partition of the square lattice \cite{zamir_shamai_erez}. See Fig.~\ref{fig:chess1} and Fig.~\ref{fig:chess2}, 
for examples. The relay message is incorporated in the user decoder to 
enhance the channel (initial) log-likelihood ratios (LLR) values. In 
other words, only initial LLR computations in the receiver decoder has 
to be modified to integrate the relay message in decoder.  

We simulate the performance of the proposed strategy using bit-interleaved 
coded modulation (BICM) and low-density parity-check codes (LDPC), 
assuming perfect channel state information at the decoder. In practice, 
channel parameters can be estimated using pilot subchannels with 
relatively small overhead, specially in femtocell scenarios with slow 
channel variations.  We study matched and  mismatched decoding metrics, 
where the decoder is aware of the interference signal alphabet (constellation), 
or treats interference as Gaussian, respectively. In the matched scenario, 
where the decoder searches over the product of user constellations 
to compute LLRs,  our results indicate that an SNR gain of approximately 
1 dB can be achieved for every bit relayed. The mismatched 
scenario is more interesting from a practical perspective, since it leads to a significantly lower decoding complexity, and our 
results  indicate that a signficant gain is obtained using the 
relay. See Fig.~\ref{fig:mber}.

The proposed relay strategy can be viewed as an analog version 
of digital network coding for wireless channels \cite{wireless_network_coding,li_yeung_cai}. We explain this 
connection to network coding and the relation of the proposed 
scheme with respect to the compress-and-forward (CF) relaying scheme \cite{cover_elgamal,lim_kim_elgamal,peyman_yu_ita10} 
in Section~\ref{sec:back}. We show that for an interference 
channel where interference is treated as noise, a scalar quantization 
strategy is asymptotically optimal. That is, we show that asymptotically 
in the low noise regime, every bit relayed improves the achievable 
sum rate by two bits, using a scalar quantize-and-forward strategy. 
This two-for-one improvement in sum rate is analogous to the rate 
improvement due to parity forwarding in the celebrated butterfly 
network example in network coding \cite{ahlswede_network_coding}.

The interference channel with an out-of-band relay has been the subject 
of a number of recent studies, where the fundamental information 
theoretic aspects of the interference relay channel are investigated 
\cite{sahin_erkip_icancellation,sahin_simeone_erkip,sahin_erkip,maric_dabora_goldsmith}. 
However, practical coding strategies to utilize relays in interference 
scenarios are not as widely studied. Moreover, not many practical 
compress-and-forward relay coding schemes are available, since  the 
decode-and-forward scheme  is often favored  upon the compress-and-forward 
relay strategy for multihop relaying. In this work, we make further 
progress in bringing some of the theoretical achievements on the application 
of relays in interference scenarios to practice  by devising a simple, 
yet effective, coding strategy based on the compress-and-forward scheme, 
tailored specifically for the interference channel. 

The rest of this paper is organized as follows: Section~\ref{sec:mot} 
describes some of the connections between the proposed strategy, 
digital network coding, and compress-and-forward relay strategy. 
Section~\ref{sec:sys} illustrates the system model and encoding 
and decoding procedures, and Section~\ref{sec:sim} provides some 
simulation results
. Finally, a few remarks conclude the paper in 
Section~\ref{sec:con}.

\section{Background \label{sec:mot}\label{sec:back}}
\subsection{Network Coding}
\begin{figure}
  \centering
  \includegraphics[width=8cm]{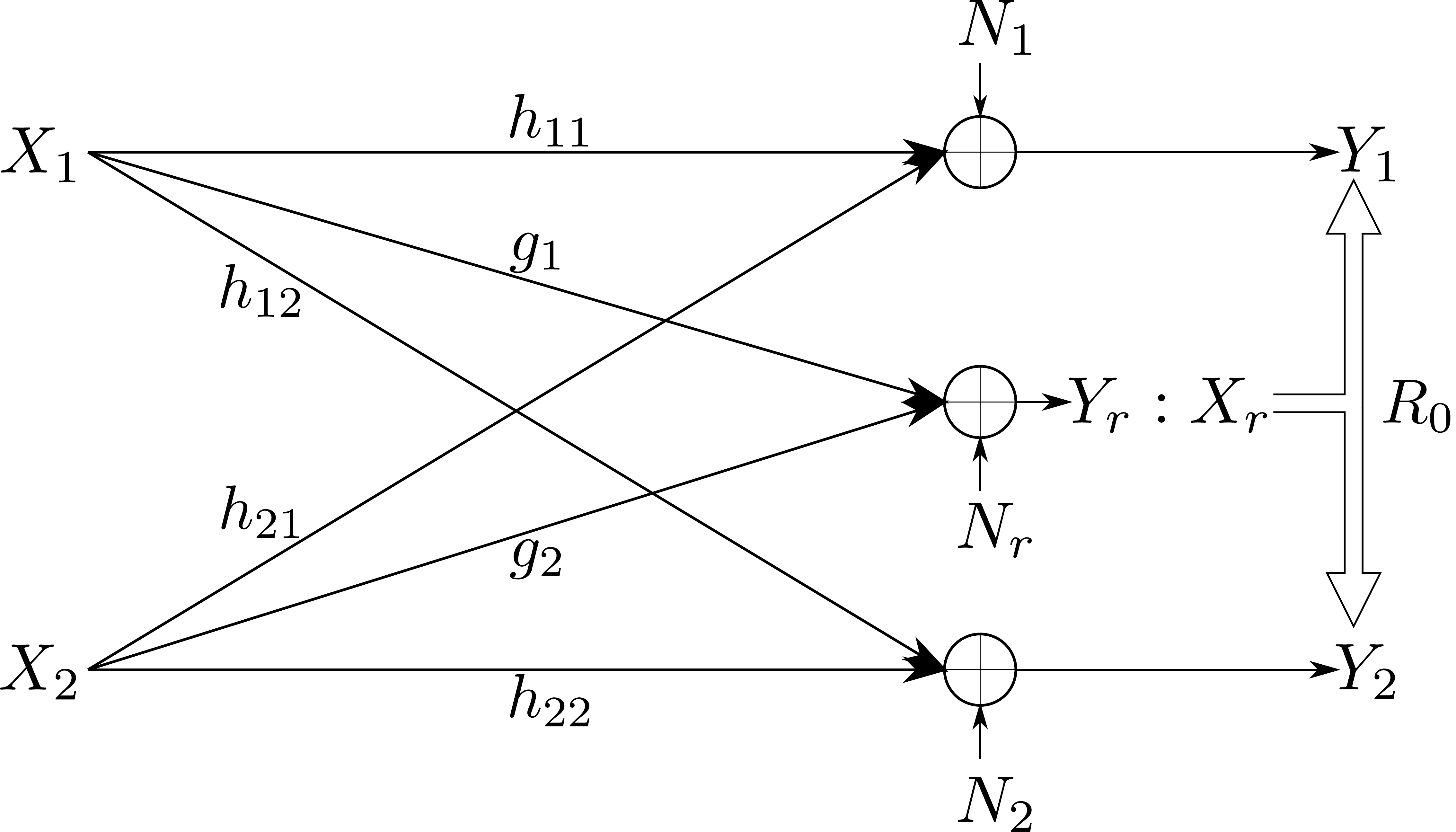}
  \caption{A two-user interference channel with a common out-of-band relay link of rate $R_0$. \label{fig:model}}
\end{figure}

\begin{figure}
\centering
\includegraphics[width=7cm]{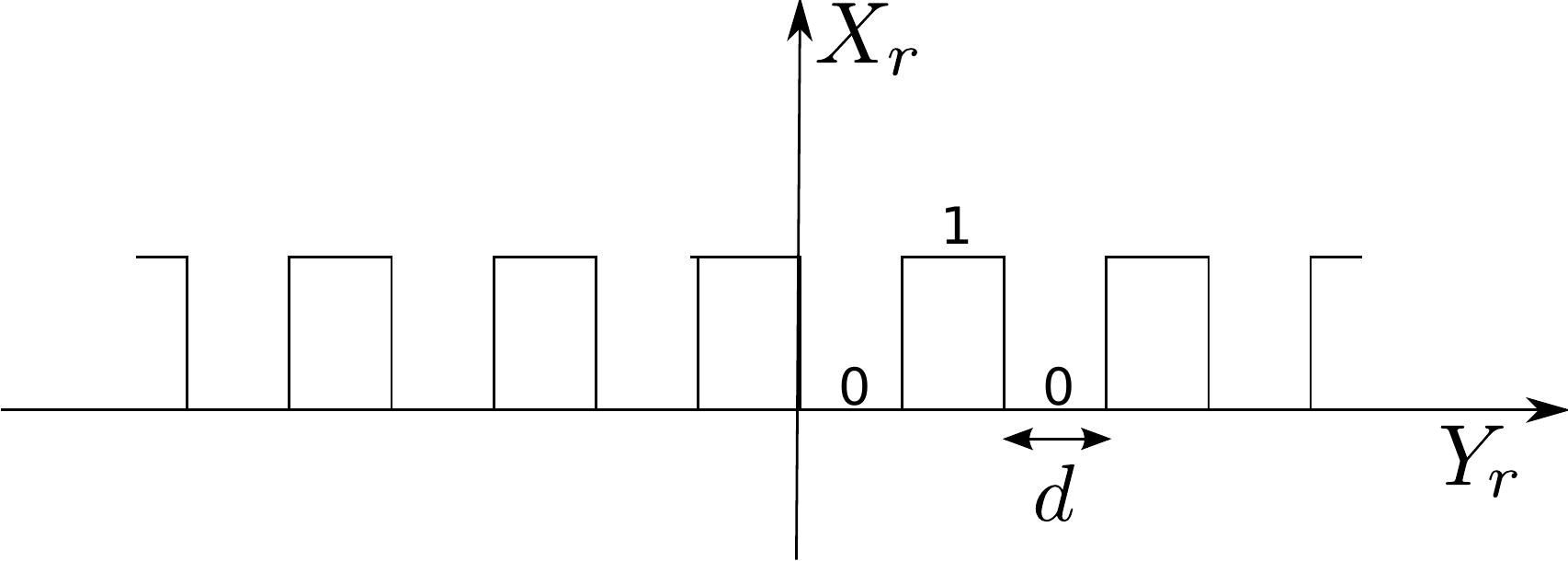}
\caption{Coarse parity: the relay forwards a parity bit for a quantized value of $Y_r$. \label{fig:toy-ill}}
\end{figure}
A simple illustration of coarse network coding can be given using a toy example. Consider a noiseless version of the interference channel shown in Fig.~\ref{fig:model} with $N_1=N_2=N_r=0$. In this model, the source signal $X_1$ is corrupted with interference $X_2$ and is received as $h_{11}X_1+h_{21}X_2$ at the destination, with $X_2$ treated as noise. Assume that all channel gains are available at destination nodes and $X_1$ and $X_2$ are real independent Gaussian random variables.  Consider now a  relay that observes $Y_r=g_{1}X_1+g_{2}X_2$ and wishes to assist the destination by forwarding 1 bit of information over a shared noiseless link.  A simple relay strategy to improve the achievable rate at the destination by close to one bit is to send a 0 if $\lfloor Y_r/d \rfloor$ is even, or a 1 otherwise, for a very small value of $d>0$; see Fig.~\ref{fig:toy-ill}. To see this, let $X_r=\lfloor Y_r/d\rfloor \mod 2$ denote the relay message, and then the rate improvement is given by:
\begin{align}
I(X_1;Y_1,X_r)&=I(X_1;Y_1)+I(X_1;X_r|Y_1)\non\\
&=I(X_1;Y_1)+I(X_1;X_r|h_{11}X_1+h_{21}X_2)\non\\
&=I(X_1;Y_1)+H(X_r|Y_1)-H(X_r|X_1,Y_1)\non\\
&\overset{(a)}=I(X_1;Y_1)+H(X_r|h_{11}X_1+h_{21}X_2)-0\non\\
&\overset{(b)}=I(X_1;Y_1)+1\non
\end{align}
In the above derivation, (a) follows since $Y_r$ is known given $X_1$ and $Y_1$ and thus,  $H(X_r|X_1,Y_1)=0$, and (b) follows since for very small values of $d$,  $\lfloor Y_r/d\rfloor \mod 2$ is a Bernoulli 1/2 random variable given $Y_1$, provided that the relay observation is not statistically identical with the receiver observations, i.e., the matrix  
\begin{align}
\left[\begin{array}{cc} h_{11} & g_1\\ h_{21} &g_2
  \end{array}
  \right] \non
\end{align}
is full rank. (See Appendix.)

Now, switch the roles of $X_1$ and $X_2$, and consider the second destination who is interested in decoding $X_2$, while $X_1$ is now interference and is treated as noise. Interestingly, the same relay strategy improves the achievable rate of the second user also by 1 bit. In other words, a single bit from the relay recovers one bit of information at each of the two destinations. This is an example of {\em coarse network coding}, where an even/odd indicator for a quantized version of relay observation recovers two bits of information, and is reminiscent of the celebrated butterfly network example in network coding \cite{ahlswede_network_coding}. A simple generalization of the above coarse parity function can be used to allow for higher relay data rates, and more importantly, to account for background noise. 

Although focusing on different problems, there are interesting connections between the proposed coarse network coding and the {\em analog network coding} strategy devised in \cite{analog_network_coding}. Analog network coding proposed in \cite{analog_network_coding} is a symbol-wise nonlinear amplify-and-forward, where a scalar relay function is optimized (assuming differentiability) such that the end-to-end  mutual information between the source and destination is maximized. Interestingly, the optimized nonlinear relay functions have a semi-periodic form, resembling a smooth version of the scalar quantization and binning of coarse network coding (see Section~\ref{sec:enc}). The analog network coding strategy, however, is devised only to serve a single destination since the relay operation depends on the destination channel, and the system optimization searches only for differentiable relay functions, which may not be necessarily optimal. 

The problem considered in this paper differs from the wireless network coding approach of \cite{wireless_network_coding}, where it is assumed that the relay observes clean versions of the source data. Coarse network coding also differs from the {\em compute-and-forward} scheme of \cite{compute_and_forward} in that structured codes at the sources are not required in coarse network coding as no decoding, even of a function of the two source messages, is performed at the relay. Further, coarse network coding is a {\em symbol-wise} strategy in the same spirit of the XOR strategy of digital network coding.

\subsection{Compress-and-Forward Relaying}
Coarse network coding is essentially a scalar compress-and-forward strategy (\cite[Theorem~6]{cover_elgamal}) where quantization and binning are performed at the symbol level. Though a scalar quantization and binning are not generally the strategies of choice for a single-relay channel, this paper shows that there are substantial gains to be obtained from a simple scalar compress-and-forward strategy in an interference channel. 


The asymptotic incremental optimality of scalar compress-and-forward shown in the previous section (i.e., one bit rate improvement per one bit relayed as noise tends to zero) is a consequence of cross-determinism in a noiseless interference channel. A cross deterministic relay channel was first introduced in \cite{cover_kim_deterministic}, and consists of a three-terminal network of a source $X$, a relay $Y_r$, and a destination $Y$, with a deterministic relationship between the relay observation $Y_r$, and the sent and received signals $X$ and $Y$, in way that $Y_r$ is a deterministic function of $X$ and $Y$. The noiseless interference channel in Section \ref{sec:mot} is an instance of two coupled cross-deterministic relay channels: here, from $X_1$ and $Y_1$ or from $X_2$ and $Y_2$, we can compute the relay observation $Y_r$; for both users, the relay channel is  cross-deterministic.

To see the role of cross-determinism in the compress-and-forward strategy consider the achievable rate of compress-and-forward with list decoding for the interference channel with interference treated as noise
\cite{peyman_yu_ita10}:
\begin{align}
R_1&\leq I(X_1;Y_1)+\min\{R_0,I(Y_r;\widehat{Y}_r|Y_1)\}-I(Y_r;\widehat{Y}_r|X_1,Y_1)\non\\R_2&\leq I(X_2;Y_2)+\min\{R_0,I(Y_r;\widehat{Y}_r|Y_2)\}-I(Y_r;\widehat{Y}_r|X_2,Y_2),\label{eq:cfrates}
\end{align}
where $\widehat{Y}_r$ is an auxiliary random variable that plays the role of quantization of the relay observation $Y_r$.
When $Y_r$ is a deterministic function of $X_1$ and $Y_1$, and also $X_2$ and $Y_2$, the penalty terms in \eqref{eq:cfrates}  vanish, since
\begin{align}
I(\hat{Y}_r;Y_r|X_1,Y_1)&=0\non\\
I(\hat{Y}_r;Y_r|X_2,Y_2)&=0\non.
\end{align}
Consequently, the rate of each user is improved by $R_0$ bits, as long as the minimums in \eqref{eq:cfrates} occur at $R_0$  for the quantization scheme used. In other words, performance is not very sensitive with respect to the quantization scheme and in particular, the block length over which quantization is performed
.

Unfortunately, optimal vector quantization with side information and list decoding as developed in \cite{lim_kim_elgamal} or \cite{peyman_yu_ita10} is  not quite amenable to practical code construction. The list decoding scheme of \cite{peyman_yu_ita10}, or the more recent noisy network coding approach of \cite{lim_kim_elgamal}, search over all possible quantized relay observations $\widehat{Y}_r$ and transmitted source codewords over a large block. In other words, an exponential number of codewords have to be examined in order to decode the source message.

However, this paper shows that much of the promised gain can be achieved at high SNRs by simply using scalar quantization and binning. This is because the requirements on $\widehat{Y}_r$ to achieve $R_0$ bits of improvement are quite loose at high SNRs, in the sense that $\widehat{Y}_r$ is admissible as long as it is sufficiently close to $Y_r$ (measured in terms of $I(Y_r;\widehat{Y}_r|Y_1)$ and $I(Y_r;\widehat{Y}_r|Y_2)$). Driven by the above intuition, we can replace the high-dimensional vector quantizer of the general CF scheme with a simple scalar quantizer. Most remarkably, when this strategy is combined with conventional (single-user)
powerful coded-modulation based on binary LDPC with BICM and iterative decoding, and the iterative decoder inputs are enhanced by incorporating the relayed bits, it is shown that a significant gain is achieved even at practical (non-asymptotically high) SNR.

\section{System Design \label{sec:sys}}
Fig.~\ref{fig:model} shows the baseband-equivalent model of a two-user interference channel with a relay defined as:
\begin{subequations}\label{eq:model}
\begin{align}
Y_1&=h_{11}X_1+h_{21}X_2+N_1\\
Y_2&=h_{22}X_2+h_{12}X_1+N_2,
\intertext{with a relay observing}
Y_r&=g_1X_1+g_2X_2+N_r,
\end{align}
\end{subequations}
where $X_1$ and $X_2$ are source signals of maximum power $P_1$ and $P_2$, and $Y_1$, $Y_2$, and $Y_r$ denote the channel observation at the two destinations and the relay, respectively. The relay assists both destinations simultaneously by broadcasting $X_r$ over a common digital link of rate $R_0$. Here, $N_1$, $N_2$, and $N_r$ are circularly-symmetric complex Gaussian noise of zero mean and variance $N_0/2$ per real dimension, and $h_{ij},g_i,i,j=1,2$ represent channel gains. We assume that the channel undergoes fast fading with independent channel gains across different time slots, and $h_{ij},g_i,i,j=1,2$ are modeled as independent circularly-symmetric complex Gaussian random variables of zero mean and variance $0.5$ per real dimension.  Also, relevant channel state information (CSI) is known to each user, specifically,  $g_1$ and $g_2$ are known to the destinations and the relay 
, and $h_{22}$, $h_{12}$ are known to destination 1 
, and  $h_{22}$,  $h_{12}$ are known to destination 2 
. In practice, the i.i.d. fast fading model with CSI known to all can be approached by using interleaving and pilot-assisted channel estimation, for a time- and frequency-selective fading channel with OFDM\footnote{Orthogonal Frequency-Division Multiplexing.},  where sufficiently long codewords are interleaved in the time-frequency domain, as currently done in today's OFDM systems. 

\subsection{Encoding: \label{sec:enc}}
The source message is encoded using a conventional LDPC code, then the binary codeword is interleaved and mapped to a sequence of M-ary constellation symbols using Gray labeling. The sequence of constellation symbols is sent over the channel using quadrature amplitude modulation (QAM). In~\eqref{eq:model}, $X_1$ and $X_2$ represent the constellation symbol sent by the two users in each time slot, and $X_r$ represent the relay message of rate $R_0$, i.e., $H(X_r)=R_0$. 

\begin{figure}
  \centering
  \includegraphics[width=6cm]{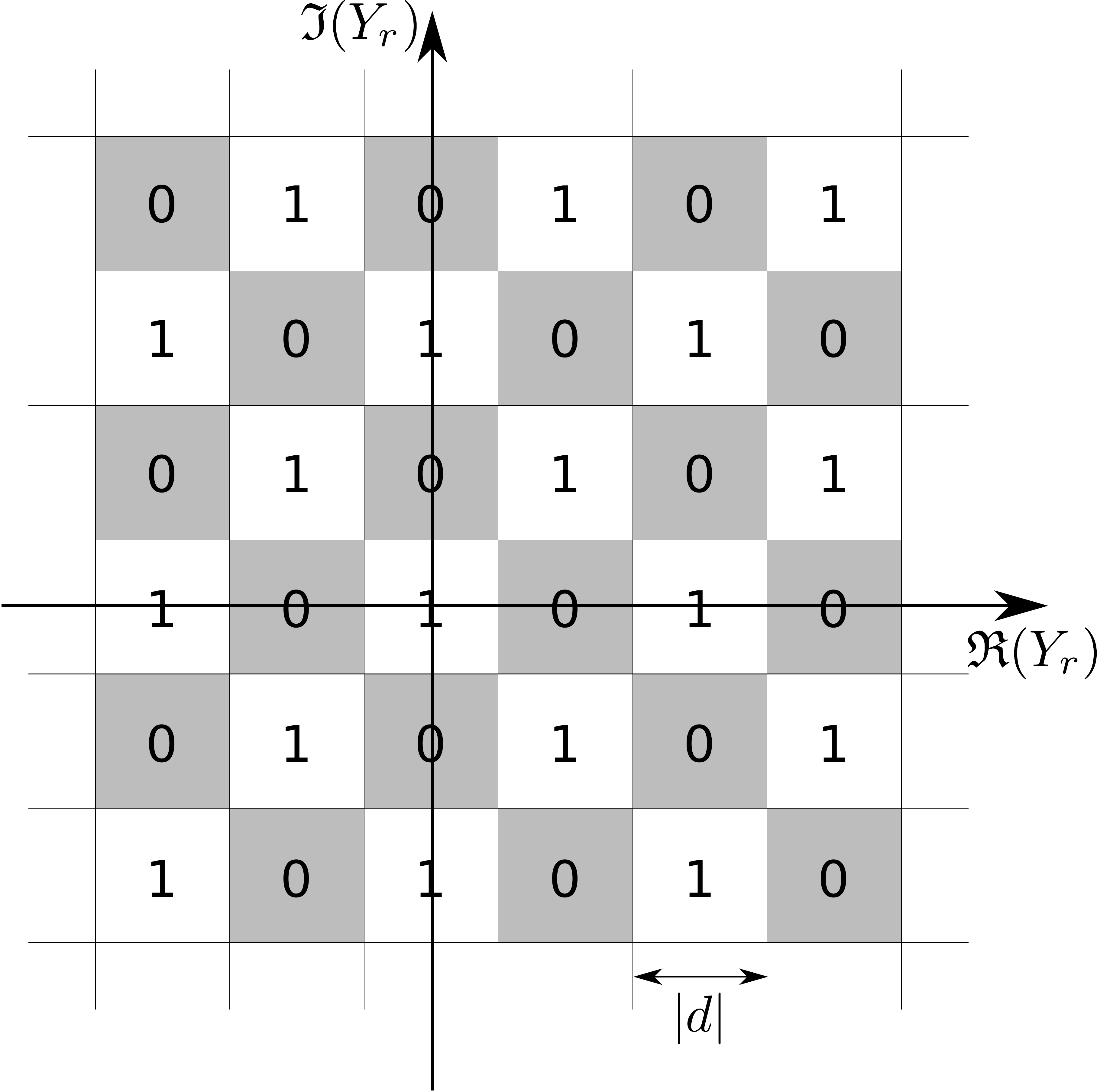}
  \caption{Chessboard quantization of $Y_r$ at rate $R_0=1$. \label{fig:chess1}}
\end{figure}

\begin{figure}
  \centering
  \includegraphics[width=6cm]{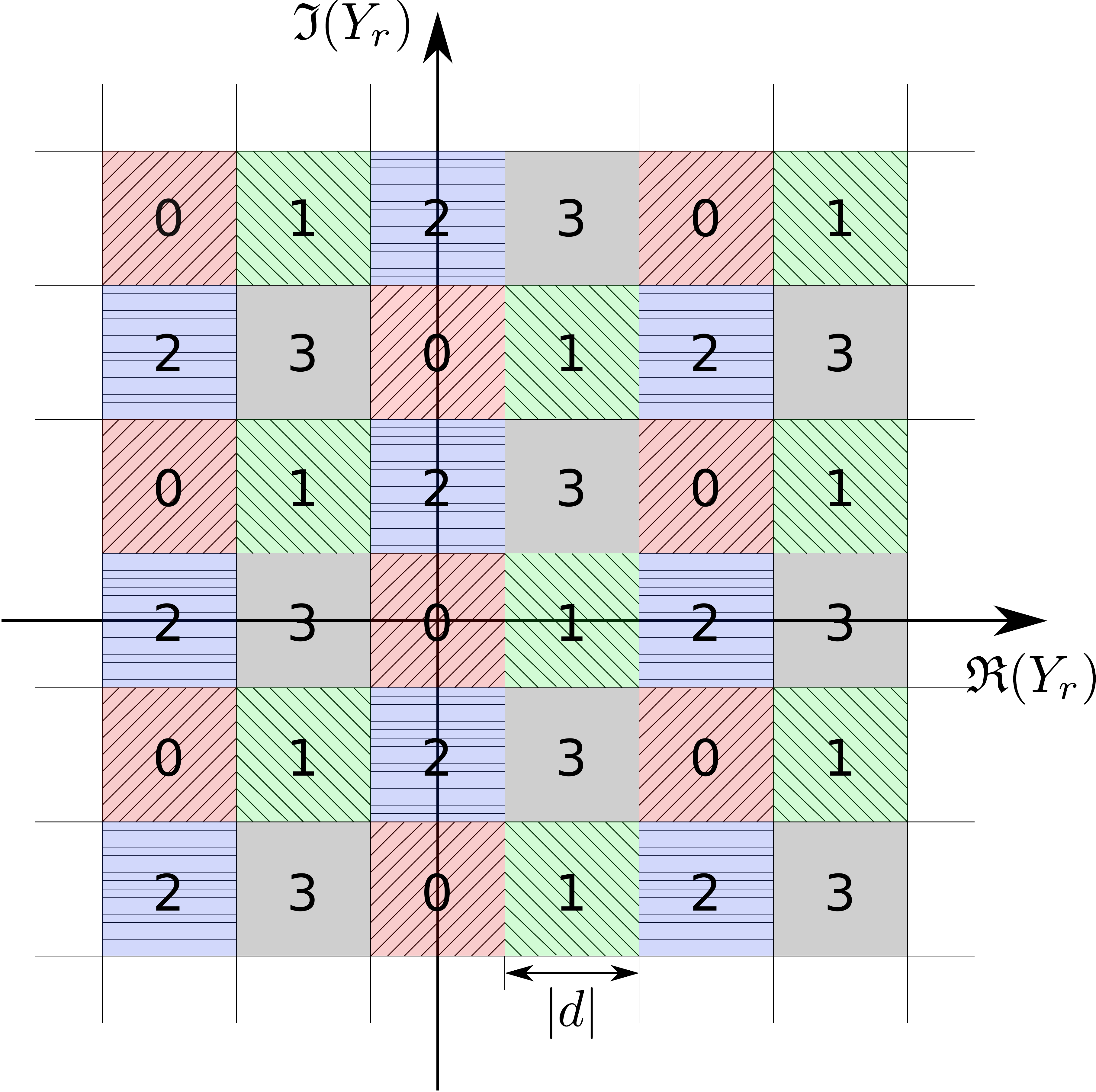}
  \caption{Chessboard quantization of $Y_r$ at rate $R_0=2$. \label{fig:chess2}}
\end{figure}

In each slot, $X_r$ is computed according to a scalar quantization scheme that we refer to as {\em chessboard quantization}, since the relay quantization Voronoi  pattern resembles a chessboard. Chessboard quantization can be described as coset partitioning of the square lattice into $2^{R_0}$ sublattices \cite{zamir_shamai_erez}. Denote the  the square integer lattice in the complex plane by $\ZZ^2$, and let $\Lambda$ denote a nested sublattice of $\ZZ^2$ with generator matrix $G$ and nesting ratio of $2^{R_0}$, i.e.,
\begin{align}
\Lambda=\{G\cdot\bi:\bi\in\ZZ^2\},
\end{align}
where $G$ is a matrix of integers with
\begin{align}
\det(G)=2^{R_0}.
\end{align}
We use $\Lambda$ to partition $\ZZ^2$. Chessboard quantization can then be formally defined as:
\begin{align}
X_r=B\left[Q(Y_r/d)\right],
\end{align}
where $d$ is a complex scaling factor, $Q(\cdot)$ represents the nearest neighbor quantizer with respect to $\ZZ^2$, and $B[~\bi~]$ denotes the coset leader for $\bi\in\ZZ^2$ with respect to the coarse lattice $\Lambda$.

For given $R_0$, $g_1,g_2$, $N_0$, and input alphabets, we need to optimize the quantization, i.e, $d$, and $\Lambda$. For practical feasibility, we restricted the quantization lattice to be $\ZZ^2$ and we have optimized $\Lambda$ independently of the channel realization, i.e., $\Lambda$ is chosen depending on the statistics of the channel coefficients and kept fixed for all channel uses.  A more complex alternative would be to choose $\Lambda$ among all sublattices with nesting ratio $2^{R_0}$ based on the value of the channel coefficients $g_1,g_2$. However, this more complex solution was not pursued here. In summary, only the scaling coefficient $d$ is optimized on a symbol-by-symbol basis, based on the realization of $g_1,g_2$. 
 In principle, the best partition for given source constellations and channel gains may be very complicated if we remove the lattice structure. Fixing the sublattice partition pattern first significantly reduces the optimization complexity.

To find a suitable lattice partition, we can optimize $\Lambda$ such that the normalized second moment of $\Lambda$ is maximized \cite{zamir_shamai_erez}. For special cases $R_0=1,2$, we choose $\Lambda$ as the lattice generated by
\begin{align}
G=\left[\begin{array}{cc}
1 & 0\\
1 & 2
\end{array},
\right]
\intertext{for $R_0=1$, and}
G=\left[\begin{array}{cc}
2 & 0\\
1 & 2
\end{array}
\right]
\end{align}
for $R_0=2$. See Fig.~\ref{fig:chess1} and Fig.~\ref{fig:chess2}.


The parameter $d$ is found such that $H(X_r|X_1,X_2)$ for given $g_1,g_2,N_0$ and input alphabets is minimized.  An explanation for choosing $H(X_r|X_1,X_2)$ as the optimization metric is given later based on log-likelihood ratios. As an alternative motivation, minimizing $H(X_r|X_1,X_2)$ is related  to maximizing the achievable rates. For a given channel realization, the rate improvement due to the relay for user one and user two is given by:
\begin{align}
\Delta R_1&=I(X_1;X_r|Y_1)=H(X_r|Y_1)-H(X_r|X_1,Y_1),\non\\
\Delta R_2&=I(X_2;X_r|Y_2)=H(X_r|Y_2)-H(X_r|X_2,Y_2).\non
\end{align}
Since the relay sends a common signal to both destinations and direct channel gains $h_{ij},i,j=1,2$ are not known at the relay, $\Delta R_1$ and $\Delta R_2$ cannot be jointly maximized at the relay for each time slot. An alternative would be to minimize $H(X_r|X_1,X_2)$ for given $g_1$ and $g_2$ (which are available at the relay), since for a given channel realization:
\begin{align}
H(X_r|X_1,Y_1)&\overset{(a)}{>}H(X_r|X_1,X_2,Y_1)\overset{(b)}{=}H(X_r|X_1,X_2)\non\\
H(X_r|X_2,Y_2)&>H(X_r|X_1,X_2,Y_2)=H(X_r|X_1,X_2)\non
\end{align}
where (a) follows from conditional entropy inequality, and (b) follows since for given channel gains, $X_r-(X_1,X_2)-Y_1$ forms a Markov chain. Thus,  $H(X_r|X_1,X_2)$ is a lower bound on both $H(X_r|X_1,Y_1)$ and $H(X_r|X_2,Y_2)$, which becomes tight at high SNR where background noise is negligible. Consequently, minimizing $H(X_r|X_1,X_2)$ asymptotically results in maximized $\Delta R_1$ and $\Delta R_2$.

\section{Decoding  \label{sec:dec}}
This section describes the decoding scheme for user one.  Decoding at user two follows similar steps. Given $Y_1$ and $X_r$, destination one computes LLRs for corresponding bit positions of the underlying LDPC code, and then conventional sum-product algorithm is used to iteratively decode the source binary codeword. In other words, $X_r$ only affects the initial computation of LLRs at the destination. 

We consider two forms of decoding: A matched scheme where the actual statistics for the interference signal is used for decoding, and a mismatched scheme where the distribution of the interference signal is approximated to be Gaussian with the same mean and variance. For QAM modulation, the matched decoder searches over the Cartesian product of the source and interference constellations to compute LLRs. Such an exhaustive search over the product constellation generally achieves a better performance at the cost of higher complexity. On the other hand, the  mismatched decoder searches only over the source constellation and approximates the LLRs by treating the interference signal (of discrete-alphabet) to be Gaussian. 

In practical settings, the interference signal constellation is often unknown to the decoder, since  the control channel of the interfering user usually needs to be decoded to obtain the encoding parameters like the constellation size. In such scenarios, the mismatch decoding scheme is an attractive strategy  with a lower decoding complexity and robust performance.

\subsection{Matched Decoder\label{sec:mdec}}
Let $b_1\cdots b_k$ denote the binary Gray label for constellation symbol $s\in\M$, where $\M$ is a constellation of size $M=2^k$. Define also the reverse mapping $b_i=A_i(s), i=1,\ldots,k$, where $b_i$ denotes the $i$'th bit of the binary label of $s$.

Given $y_1$ and $x_r$, the destination computes $\lambda_i$, the LLR corresponding to the $i$'th bit position, for $i=1,\ldots,k$ as follows:
\begin{align}
\lambda_i=\log\frac{\sum_{\{x_1\in\M_1,x_2\in\M_2\vert A_i(x_1)=1\}}p(y_1,x_r\vert x_1,x_2)}{\sum_{\{x'_1\in\M_1,x'_2\in\M_2\vert A_i(x'_1)=0\}}p(y_1,x_r\vert x'_1,x'_2)}\label{eq:llr1},
\end{align}
where $\M_1,\M_2$ are the constellations for user one and two.
Now, to compute $p(y_1,x_r\vert x_1,x_2)$, we have:
\begin{align}
p(y_1,x_r\vert x_1,x_2)&=p(y_1\vert x_1,x_2)p(x_r|x_1,x_2)\label{eq:llr2},
\end{align}
where
\begin{align}
p(y_1\vert x_1,x_2)&=\frac{1}{\pi N_0}\exp\left\{-\frac{\norm{y_1-h_{11}x_1-h_{21}x_2}^2}{N_0}\right\}, \non
\end{align}
and
\begin{align}
p(x_r|x_1,x_2)&=\frac{1}{\pi N_0}\iint\limits_{y_r\in \B(x_r)}\exp\left\{-\frac{\norm{y_r-g_1x_1-g_2x_2}^2}{N_0}\right\}dy_r\non
\end{align}
where $\B(x_r)$ encompasses all quantization regions with $X_r=x_r$; for example,  $\B(0)$ represents all gray-colored squares corresponding to $x_r=0$ in Fig.~\ref{fig:chess1}.

The LLR functions in \eqref{eq:llr1} and \eqref{eq:llr2} also suggest that minimizing $H(X_r|X_1,X_2)$ amounts to better initial LLRs with larger  magnitude. This can be illustrated by \eqref{eq:llr2}, where the effect of the relay signal $X_r$ on the destination LLRs is controlled by $p(x_r|x_1,x_2)$. Hence, the more {\em deterministic} $p(x_r|x_1,x_2)$ is, the larger the magnitude of initial LLRs are, which results in a lower decoding error rate.

\subsection{Mismatched Decoder}
Computing the LLR values in \eqref{eq:llr1} requires summing over all pairs of symbols in $\M_1$ and $\M_2$, which can be computationally intensive depending on the size of $\M_1$ and $\M_2$. This section describes a low-complexity mismatched decoding scheme that does not require the knowledge of  $\M_2$.    

The mismatched decoder treats the interference signal $X_2$ as a Gaussian random variable with the same mean and variance. 
Yet,  the mismatched scheme becomes matched when $X_2$ is indeed Gaussian; for example, asynchronous cross sub-channel interference in an  OFDM system due to misaligned timing can be well approximated to be Gaussian.  

Decoding is again performed by computing the  initial mismatched log-likelihood ratios (mLLR). Given $y_1$ and $x_r$, user one computes $\gamma_i$, the mLLR corresponding to the $i$'th bit position, for $i=1,\ldots,k$ as follows:
\begin{align}
\gamma_i=\log\frac{\sum_{\{x_1\in\M_1\vert A_i(x_1)=1\}}q(y_1,x_r\vert x_1)}{\sum_{\{x'_1\in\M_1\vert A_i(x'_1)=0\}}q(y_1,x_r\vert x'_1)}\label{eq:mllr1},
\end{align}
where $\M_1$ is again the constellations for user one. The metric $q(y_1,x_r|x_1)$ is given by:
\begin{align}
q(y_1,x_r|x_1)&=q_1(y_1|x_1)q_r(x_r|x_1,y_1),\label{eq:qmetric}
\end{align}
where
\begin{align}
q_1(y_1|x_1)&=\frac{1}{\pi(\norm{ h_{21} }^2 P_2+N_0)}\exp\left\{-\frac{\norm{y_1-h_{11}x_1}^2}{\norm{ h_{21} }^2 P_2+N_0}\right\}\non,
\end{align}
and
\begin{align}
q_r(x_r|x_1,y_1)&=\iint\limits_{y_r\in \B(x_r)}
\frac{1}{\pi\sigma}\exp\left\{-\frac{\norm{y_r-\mu}^2}{\sigma^2}\right\}dy_r,
\end{align}
with
\begin{align}
\sigma^2&=N_0+\frac{\norm{g_2}^2P_2N_0}{\norm{h_{21}}^2P_2+N_0}\non\\
\mu&=g_{1}x_1+\frac{g_2h_{21}^*P_2}{\norm{h_{21}}^2P_2+N_0}(y_1-h_{11}x_1).\non
\end{align}
Note that $q(y_1,x_r|x_r)=p(y_1,x_r|x_1)$ when $X_2$ is Gaussian. The mLLRs computed for every bit position are used as initial LLRs fed to the  receiver's iterative decoder.

\section{Achievable Rates}
Under the matched decoding scenario, the achievable rates of the above strategy can be computed using standard mutual information analysis. However, characterizing the capacity limits under the mismatched decoding scenario requires a different set of techniques. Based on the analysis in this section, numeric results for the achievable rates under matched and mismatched scenarios are presented in Section~\ref{sec:sim}.

\subsection{Matched Metric}
In the matched scenario,  the decoder uses $p(y_i,x_r|x_i)$ as the likelihood  metric, $i=1,2$ for user one and two. For this decoding metric, the  achievable rate  is  given by the mutual information between channel input $X_i$ and the channel outputs $Y_i,X_r$ for $i=1,2$. Thus, we have
\begin{align}
R_1&=I(X_1;Y_1X_r)=\Ex \left[ \log\frac{p(X_1,Y_1,X_r)}{p(X_1)p(Y_1,X_r)}\right]\label{eq:r1}\\
R_2&=I(X_2;Y_2X_r)=\Ex\left[ \log\frac{p(X_2,Y_2,X_r)}{p(X_2)p(Y_2,X_r)}\right]\label{eq:r2},
\end{align}
where the expectation $\Ex$ is also with respect to channel coefficients\footnote{All the probability distributions and mutual information expressions in this section are conditioned with respect to the channel coefficients, but  explicit conditioning is dropped for notational simplicity. For example, $p(y_i,x_r|x_i)$ should be treated as $p_{Y_i,X_r|X_i,h_{11},\ldots,g_1,g_2}(y_i,x_r|x_i,h_{11},\ldots,h_{22},g_1,g_2)$ and $I(X_1;Y_1X_r)$ denotes $\Ex_{h_{11},\ldots,h_{22},g_1,g_2}I(X_1;Y_1X_r|h_{11},\ldots,h_{22},g_1,g_2)$
.}. Using \eqref{eq:r1}, $R_1$ can be computed for discrete-alphabet $X_1$ and $X_2$ inputs using
\begin{align}
p(y_1,x_1,x_r)=\sum_{x_2\in \M_2}p(y_1,x_r|x_1,x_2)p(x_1)p(x_2),\label{eq:prb1}
\end{align} 
along with \eqref{eq:llr2}, where $p(x_1)$ and $p(x_2)$ are uniform over the source constellations. When $X_2$ is continuous, an equivalent expression can be found for $p(y_1,x_1,x_r)$ by using integral in place of sum in \eqref{eq:prb1}. The rate $R_2$ can be computed similarly.

\subsection{Mismatched Metric}
The mismatched decoder uses $q(y_i,x_r|x_i)$ as the likelihood  metric. The supremum of  achievable rates of a mismatched decoder is unknown in general. However, error exponent analysis with random coding can be used to establish an achievable rate for a mismatched decoder \cite{gmi}. 

\begin{theorem}[Generalized Mutual Information \cite{gmi}]\label{thm:gmi}
Consider a point-to-point channel defined by $p(y|x)$ for $X\in \X$ and $Y\in\mathcal{Y}$. Using a mismatched decoder with mismatched likelihood metric $q(x,y)$,  all positive rates $R$ are achievable for this channel with a vanishing error probability for large block lengths if
\begin{align}
R<I^{gmi}(X;Y)\triangleq\max_{s>0}I_s^{gmi}(X;Y), 
\end{align} 
where 
\begin{align}
I_s^{gmi}\triangleq \Ex\left[\log\frac{q(X,Y)^s}{\sum_{x'\in\X}p(x')q(x',Y)^s}\right].
\end{align}
\end{theorem}
Using Theorem \ref{thm:gmi}, computing lower bounds for the achievable capacity with the mismatched decoding metric in \eqref{eq:qmetric} is quite straightforward.  For user one, we have
\begin{align}
R_1(s)=\Ex\left[\log\frac{q(Y_1,X_r|X_1)^s}{\sum_{x_1\in\M_1}\frac{1}{M_1}q(Y_1,X_r|x_1)^s}\right],
\end{align}
where $M_1$ is the size of $\M_1$. The expectation $\Ex$ is with respect to  probability distribution $p(y_1,x_1,x_r)$ given in \eqref{eq:prb1}, and also the channel coefficients. To compute $R_1(s)$ numerically with fading coefficients, the average $I_s^{gmi}$ is computed for different realizations of the channel coefficients, with $s$ optimized for each realization.  A similar expression can be found for user two.

\section{Simulation Results \label{sec:sim}}
This section first investigates the theoretical achievable rate for the matched and mismatched scenarios. Then, the performance of iterative decoding for LDPC codes with BICM modulation is studied using simulations.

\subsection{Achievable Rates}
\ctable[
   cap     = {SNR Improvement},
   caption = {Improvement in Minimum required SNR to achieve $R_1=1.5$.},
   label   = {tb:gaps},
]{cccc}{
}{                                   \FL
 $\M_1$ & $\M_2$ &  $R_0=1$ & $R_0=2$    \ML
 4-QAM & 4-QAM &  0.86 dB &  2.75dB  \NN
 4-QAM & 16-QAM  &  2.03 dB &   4.74 dB\NN
 16-QAM & 16-QAM  &  2.18 dB &  4.78 dB \LL
}

For the matched decoding scenario, Table~\ref{tb:gaps} lists the improvement in minimum SNR required to achieve a source rate of 1 bit/symbol for different constellation sizes as channel inputs. As it is shown, when both the desired signal and interference are taken from 4-QAM constellations, an SNR gain of 0.86 dB is obtained with 1 bit relayed. With 2 bits relayed, the SNR gain is more substantial: 2.75 dB improvement to achieve a data rate of 1 bit/symbol. The gain due to coarse network coding improves at higher bit-per-symbol rates. The second and third row of Table~\ref{tb:gaps} list the gain in minimum SNR required to achieve $R_1=1$, when interference $X_2$ is taken from a 16-QAM constellation. In this case, 1 bit of relaying results in about 2 dB gain in SNR, and the SNR gain is almost doubled in dB with 2 bits relayed.

\begin{figure}
\centering
\includegraphics[height=7cm]{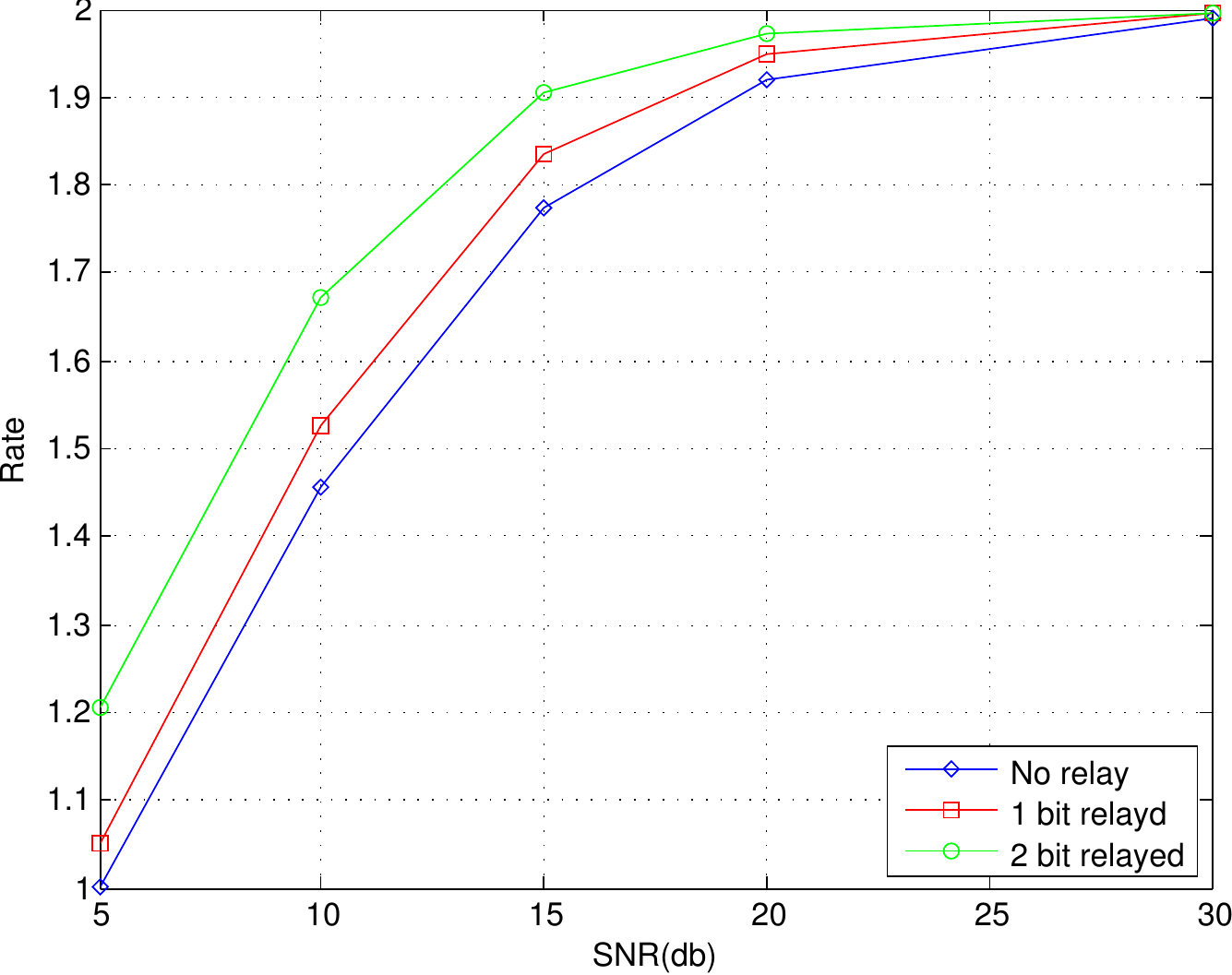}
\caption{  Achievable rate with matched decoding and  QPSK  $X_1$ and $X_2$. \label{fig:4}}
\end{figure}

\begin{figure}
\centering
\includegraphics[height=7cm]{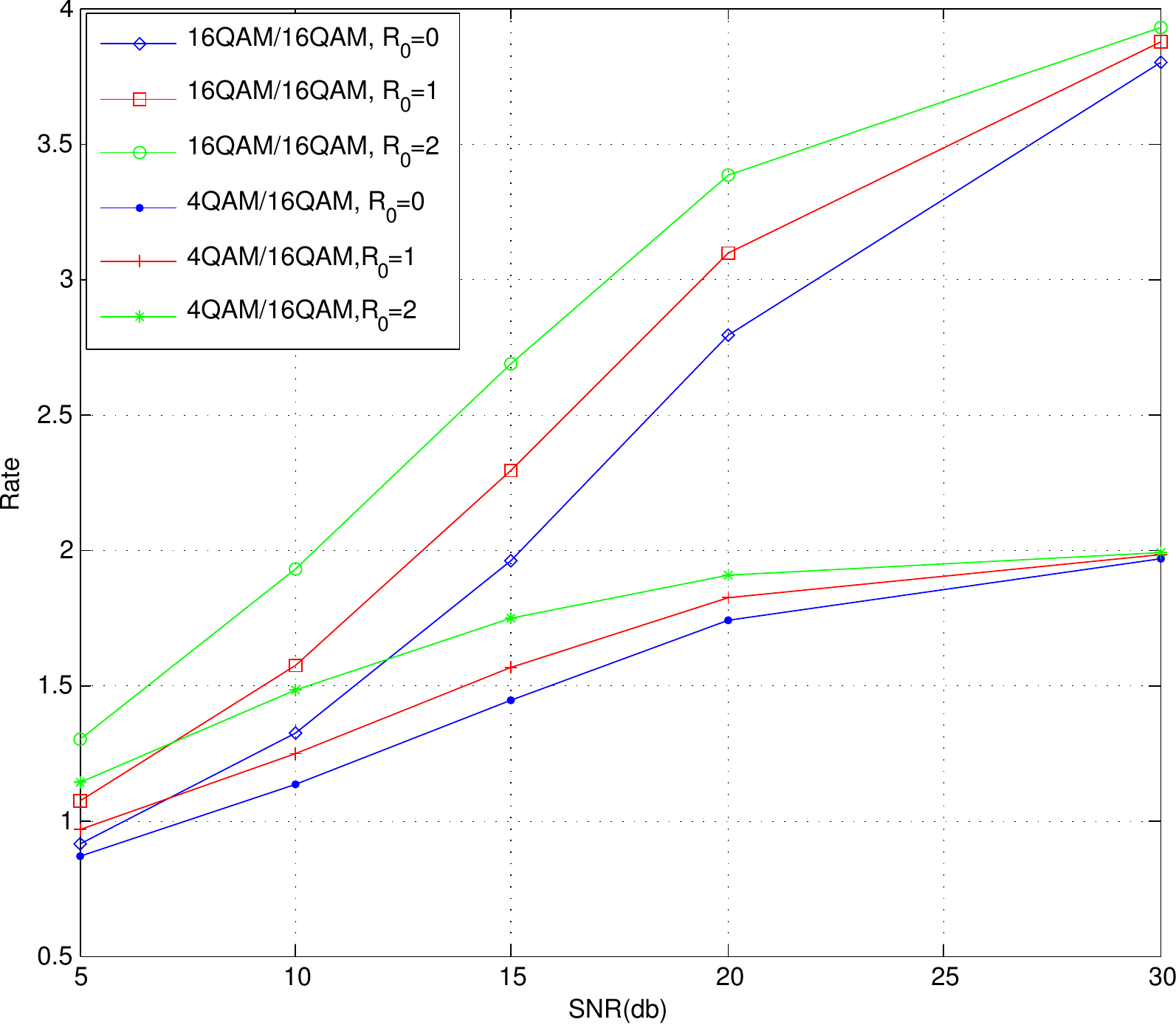}
\caption{  Achievable rate with matched decoding and 16-QAM $X_2$. \label{fig:16}}
\end{figure}

\begin{figure}
  \centering
  \includegraphics[height=7cm]{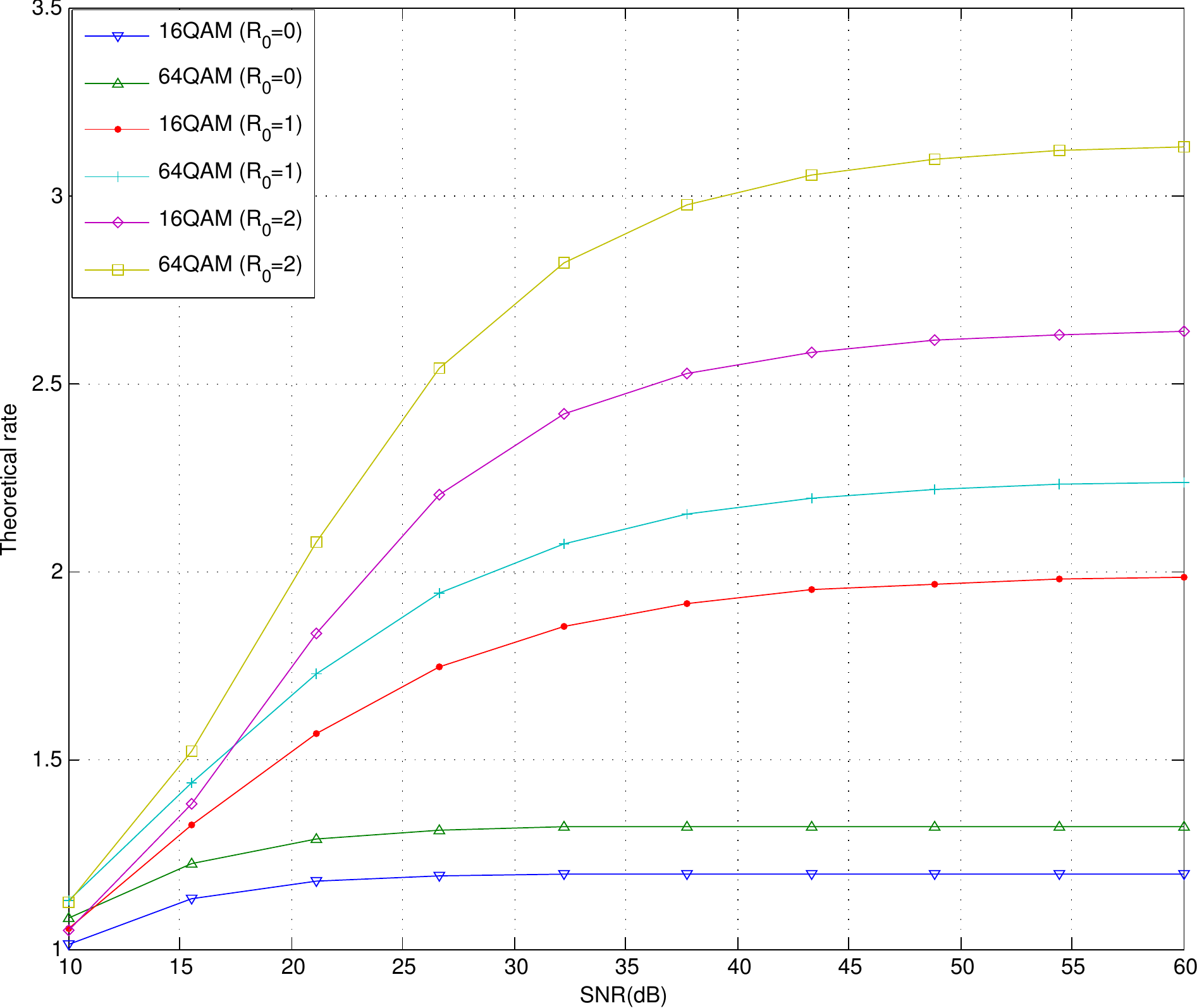}
  \caption{ Achievable rate  with matched decoding for discrete-alphabet $X_1$  and Gaussian interference $X_2$.\label{fig:gauss-inter}}
\end{figure}

Fig.~\ref{fig:4} shows the achievable rate $R_1$ when both $X_1$ and $X_2$ are taken from a 4-QAM (QPSK) constellation of power $P$, and SNR is defined as $10\log_{10}(P/N_0)$. With discrete input alphabets, the destination can uniquely identify the source symbol sent by $X_1$ at high SNRs, and thus asymptotically, the achievable rate $R_1$ tends to $\log_2(4)=2$, as shown in Fig.~\ref{fig:4}. Fig.~\ref{fig:4} also indicates that a substantial gain in minimum SNR required to achieve a certain rate is achieved. Highest SNR gains are obtained for larger values of $R_1$. At SNR=25 dB, the SNR gain is close to 4 dB for one bit relayed, and 8 dB for 2 bits relayed.


Fig.~\ref{fig:16} shows the achievable rate curves when interference comes from  a 16-QAM constellation. In this figure,  $X_1$ and $X_2$ are of the same power $P$, and SNR is $10\log_{10}(P/N_0)$. Comparing the first and second row in Table~\ref{tb:gaps}, it is revealed that coarse network coding is more helpful for a QPSK source when interference $X_2$ comes from a 16-QAM constellation as opposed to 4-QAM.
A possible conclusion is that the more damaging the interfering signal is, the more the relay can help. 

Fig.~\ref{fig:gauss-inter} shows the rate improvement obtained by coarse network coding when the source signals $X_1$ is taken from a 16-QAM and 64-QAM constellations and interference is a circularly-symmetric complex Gaussian random variable. Both $X_1$ and $X_2$ are of the same power $P$ and SNR is again  $10\log_{10}(P/N_0)$. In this case, coarse network coding almost linearly improves the achievable rate. Highest gains are achieved when $X_1$ is taken from a 64-QAM constellation,  where asymptotically at high SNRs every relayed bit improves the achievable rate by slightly less than 1 bit. 

\begin{figure}
  \centering
  \includegraphics[height=7cm]{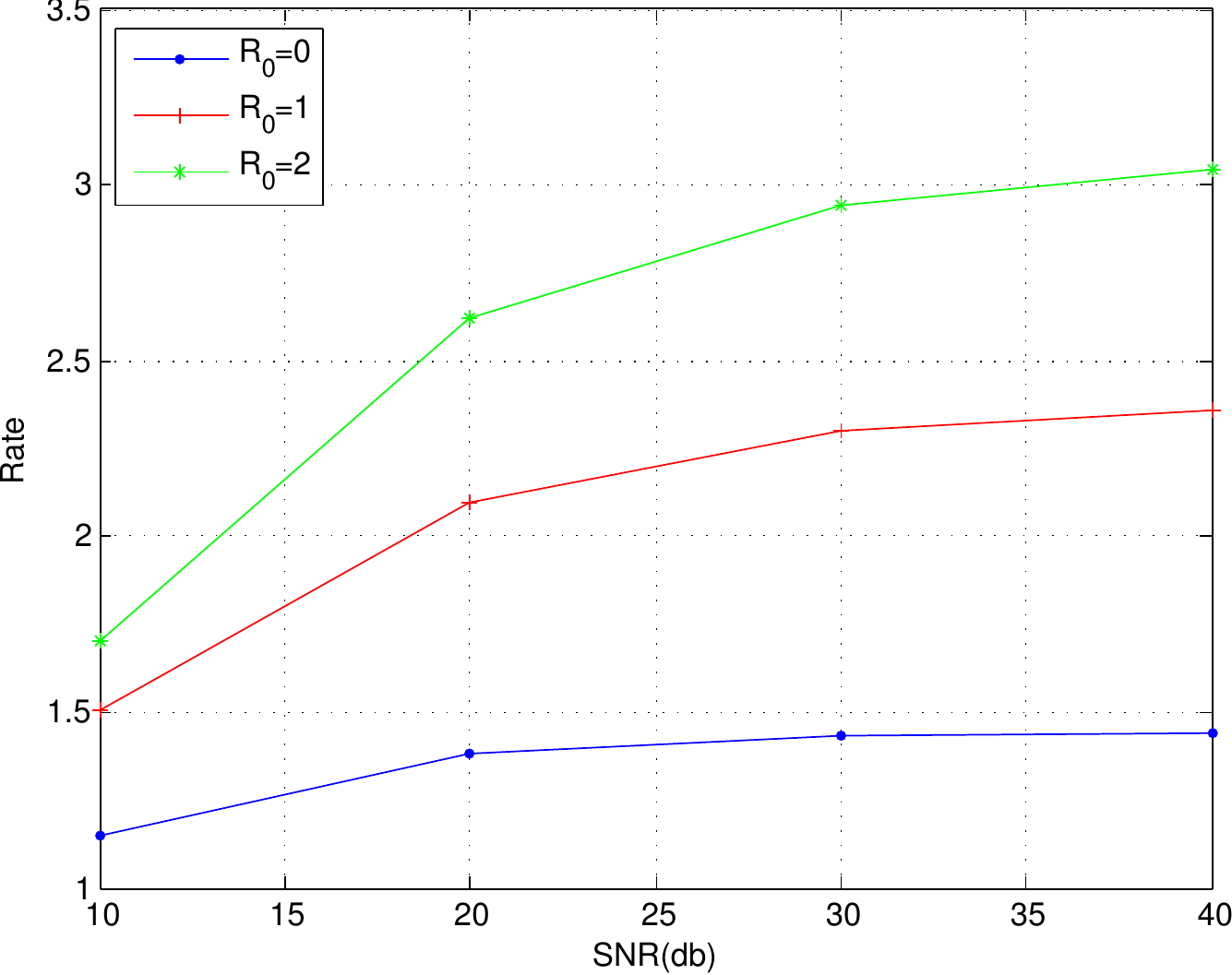}
  \caption{ Achievable rate with matched decoding for  Gaussian inputs. \label{fig:gauss}}
\end{figure}

From Fig.~\ref{fig:4}, Fig.~\ref{fig:16}, and Table~\ref{tb:gaps}, it is expected that coarse network coding improves the achievable rates by  larger extents as the size of the source alphabets increase. Fig.~\ref{fig:gauss} shows the achievable rates when both $X_1$ and $X_2$ are circularly-symmetric Gaussian random variables. It is shown in Fig.~\ref{fig:gauss} that for Gaussian inputs, the achievable rate of both users is improved by close to 1 bits at $R_0=1$, and 1.5 bits at $R_0=2$, at high SNRs. 
\begin{figure}
  \centering
  \includegraphics[height=7cm]{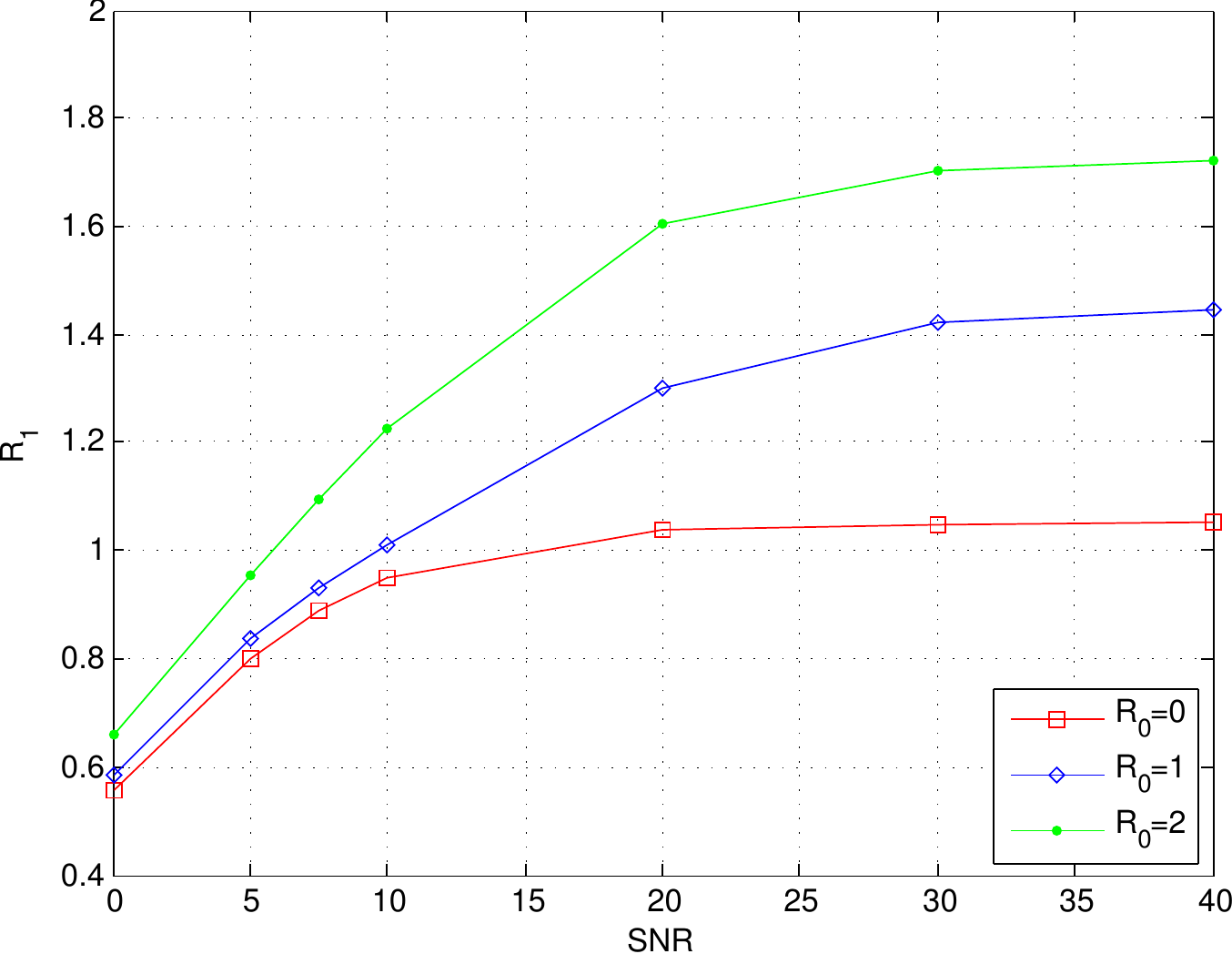}
  \caption{Achievable rate with mismatched decoding and QPSK inputs. \label{fig:mismatched-rate}}
\end{figure}

Fig.~\ref{fig:mismatched-rate} shows the achievable rate obtained by the generalized mutual information analysis for the mismatched decoding metric and QPSK inputs.  Although both the inputs $X_1$ and $X_2$ are discrete, the achievable rate with mismatched decoding is saturated even at high SNR values due to interference. Yet, adding the relay alleviates the rate saturation at high SNRs by improving the asymptotic achievable rate in proportion to the number of bits relayed. The improvement in achievable rates for QPSK inputs and mismatched decoding is less than a bit for every bit relayed. However, this rate improvement tends to one as the size of signal constellations increases. At the extreme case where both $X_1$ and $X_2$ are continuous Gaussian random variables, the mismatched decoding reduces to the matched strategy, and every bit relayed results in two bits improvement in sum rate for large SNR.

\subsection{Performance with LDPC-Encoded BICM}
\begin{figure}
  \centering
  \includegraphics[height=7cm]{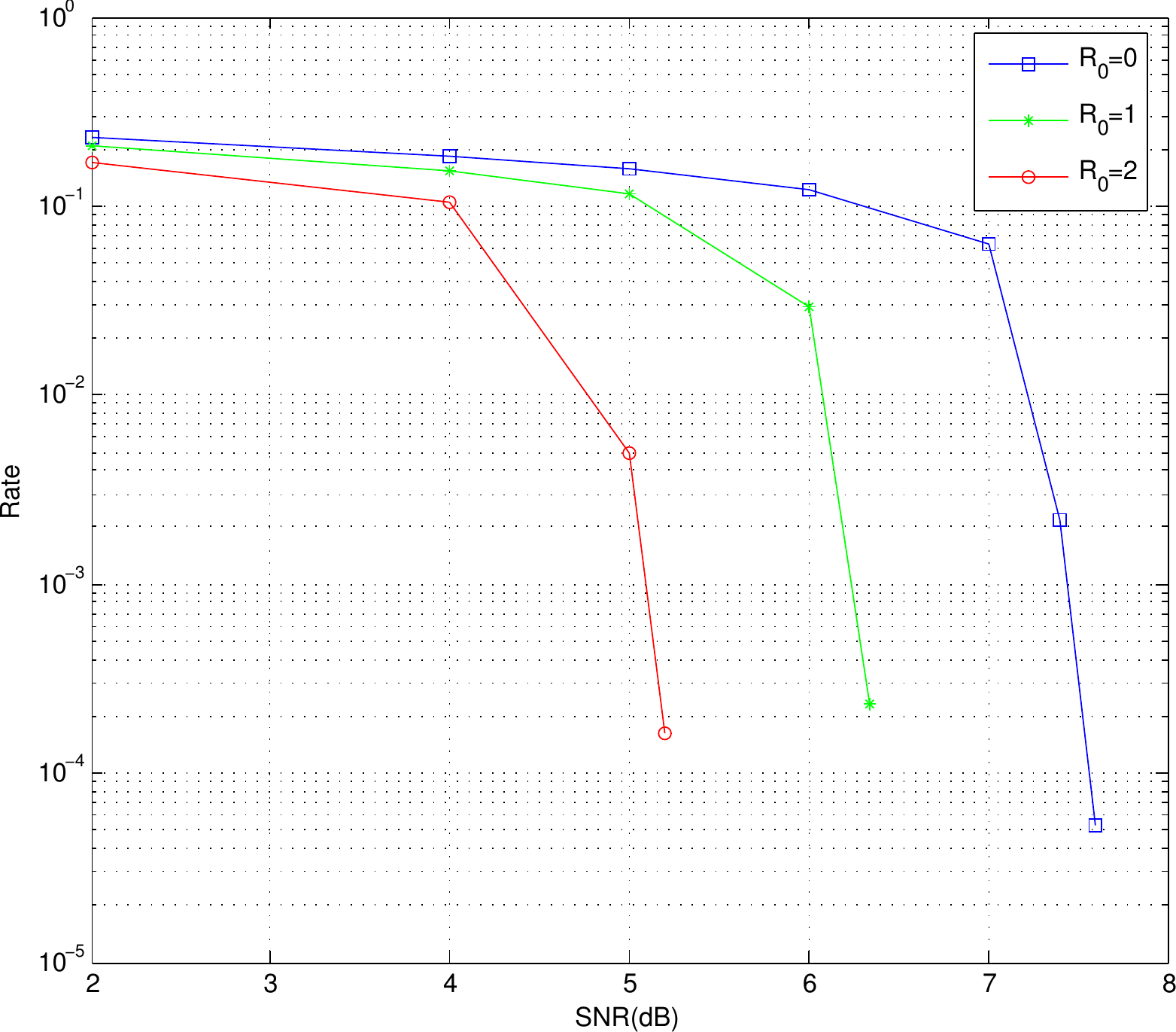}
  \caption{The bit-error-rate performance of regular (3,6) LDPC code with matched decoding and BICM gray-mapped signaling. Block length is 20000. \label{fig:ber}}
\end{figure}

For the matched scenario, Fig.~\ref{fig:ber} shows the performance of BICM-LDPC coding along with coarse network coding and iterative decoding at source rate $R_1=1$. The BICM system is comprised of an LDPC code of rate 0.5 and block length 20,000, and gray-labeled 4-QAM modulation for both $X_1$, and interference $X_2$. The underlying LDPC code is the Gallager regular (3,6) LDPC code \cite{gallager_ldpc}, with a random graph construction. Decoding is performed as described in Section~\ref{sec:dec} by enhancing the initial LLRs using the extra bit received from the relay.

An SNR gain of approximately 1 dB is achieved using the (3,6) LDPC code of rate $1/2$  along with coarse network coding at $R_0=1$. At  $R_0=2$, the gain in SNR is approximately close to 2.5 dB. Improved performance could perhaps be achieved by optimizing LDPC codes for the specific enhanced LLR density distribution. Because of the decoding approach used, the achieved gain is controlled by the shape of enhanced LLR density distribution. The more the relay message skews the initial LLR density distribution, the higher the achieved SNR gain would be, which also requires tuning the LDPC code for the specific input LLR density distribution to fully exploit potential gains.  In this paper, standard LDPC codes are used to simulate the performance,  as our focus is more on the multiuser aspects, rather than the impact of LDPC graph itself. Also, since the relay message is only used for initial LLR computations, the iterative decoding procedure at each destination is equivalent to that of a single-user point-to-point channel, which can be optimized by standard tools such as density evolution, EXIT charts, and protograph-based structures \cite{ldpc-book,shokrollahi_urbanke,chung_forney_001db}.

\begin{figure}
  \centering
  \includegraphics[height=7cm]{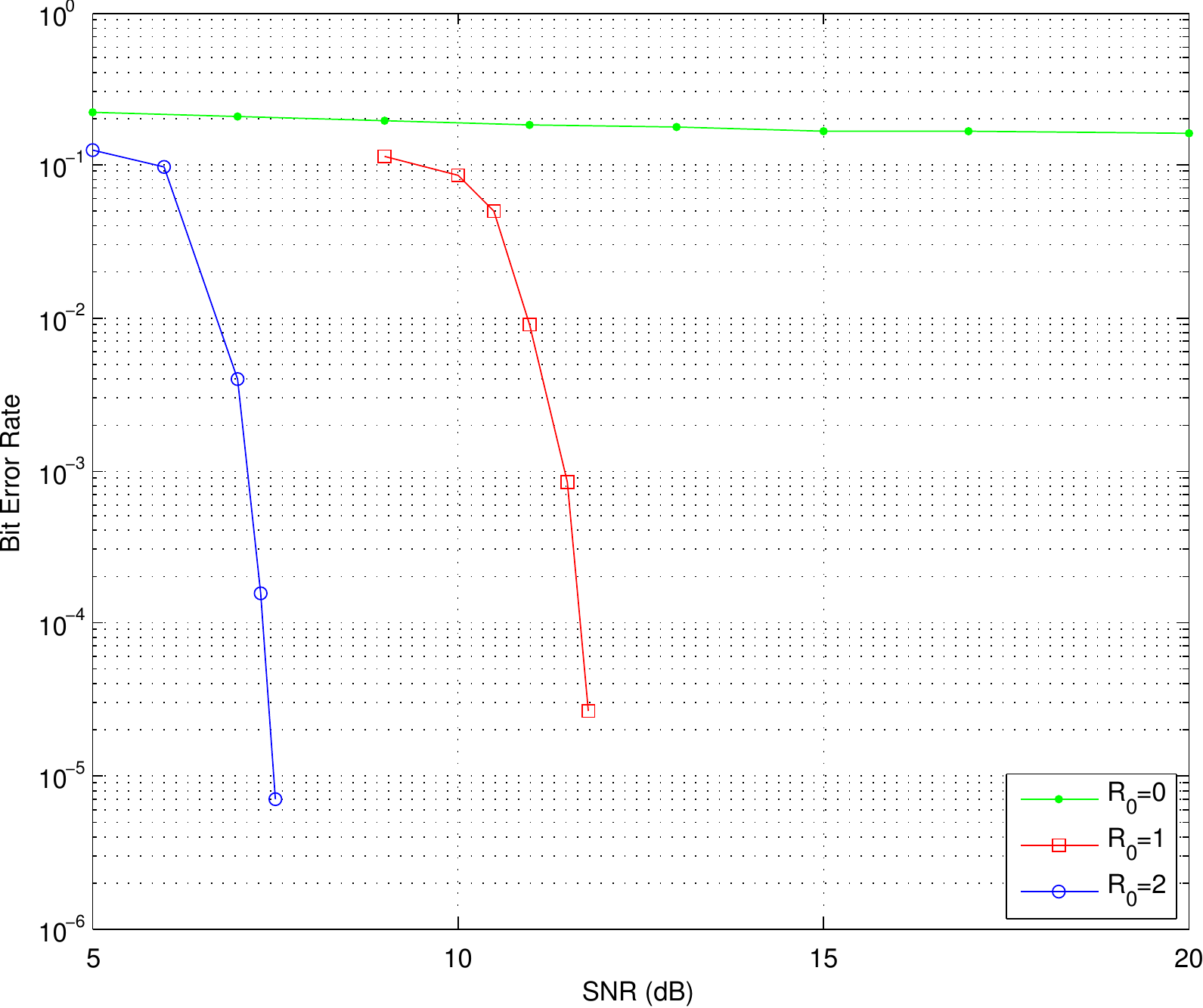}
  \caption{The bit-error-rate performance of regular (3,6) LDPC code with mismatched decoding and BICM gray-mapped signaling. Block length is 20000. \label{fig:mber}}
\end{figure}

Fig.~\ref{fig:mber} shows the performance of the (3,6) LDPC code of
rate 0.5 with mismatched decoding and gray-labeled QPSK
BICM signaling. Using the relay results in substantial im-
provements when the decoder is unaware of the interfering
signal constellation. At rate $R_1 = 1$, receiver is unable to
decode the source codeword with mismatched LLR values,
even at high SNRs, without relay assistance. Yet, using a relay
link of rate $R_0 = 1$ enhances the LLR values in a way that
decoding is possible at SNR values around 10 dB, which is
consistent with GMI achievable rates in Fig. 11. A relay link
of rate $R_0 = 2$ results in an additional 5 dB SNR gain as
compared to $R_0 = 1$, again matching the result from GMI simulation in Fig.~\ref{fig:mismatched-rate}. The  SNR gains are quite substantial, especially if we consider the very simple relay strategy, requiring no decoding and only scalar quantization.

\subsection{An intuitive interpretation}
It is helpful to intuitively understand how \eqref{eq:llr1} and \eqref{eq:qmetric} enhance the LLR values at the decoder. We can gain some insights by looking at likelihood computation for a particular constellation symbol $s_1$ sent by user one. Assuming that $s_1$ is sent by user one, the decoder can subtract $s_1$ from its observation $y_1$ to find an estimate for $s_2$, the symbol sent by user two. Now, using the channel state information, decoder can find an estimate for the relay observation and the corresponding quantization index. If the decoders estimate of the relay message matches the message received from the relay, the likelihood of $s_1$ being sent is improved, otherwise, it is less likely that $s_1$ is sent by user one. Such small adjustments to the LLR values at symbol level can be combined over a larger block length  using, for example, an LDPC code with iterative decoding. Small enhancements in initial LLR values translate to significant improvement in overall bit error rate (BER) performance, because of  the 
waterfall characteristic of BER curves for LDPC codes.


\section{Conclusion \label{sec:con} \label{sec:rcoop}}
Interference is one of the main obstacles to deploy user-based uncoordinated  wireless cellular networks. We propose a simple relay strategy to improve the achievable rates in a two-user interference channel, where interference is treated as noise. This relay strategy is called coarse network coding, since the relay operation involves a scalar quantization followed by a scalar binning, reminiscent of the parity generation in digital network coding. In addition, asymptotically at high SNRs, a single bit sent by the relay in coarse network coding improves the rate of both users by one bit, resembling the rate improvement obtained from a single parity bit at two different destinations in digital network coding. In this relay scheme, the relay quantizes its observation using a scalar quantization obtained by suitable partitioning of the square lattice. We consider decoding scenarios using matched and mismatched metrics, depending on whether where the alphabet of interfering signal is known or unknown to the decoder, respectively. Using LDPC-encoded BICM modulation, we show that the expected theoretical SNR gains due to coarse network coding are mostly achieved using a generic regular (3,6) LDPC code; higher SNR gains are expected by designing the LDPC graph for the specific density distribution of the relay-enhanced LLRs. This is left for future work.

\begin{figure}
\centering
\includegraphics[width=8cm]{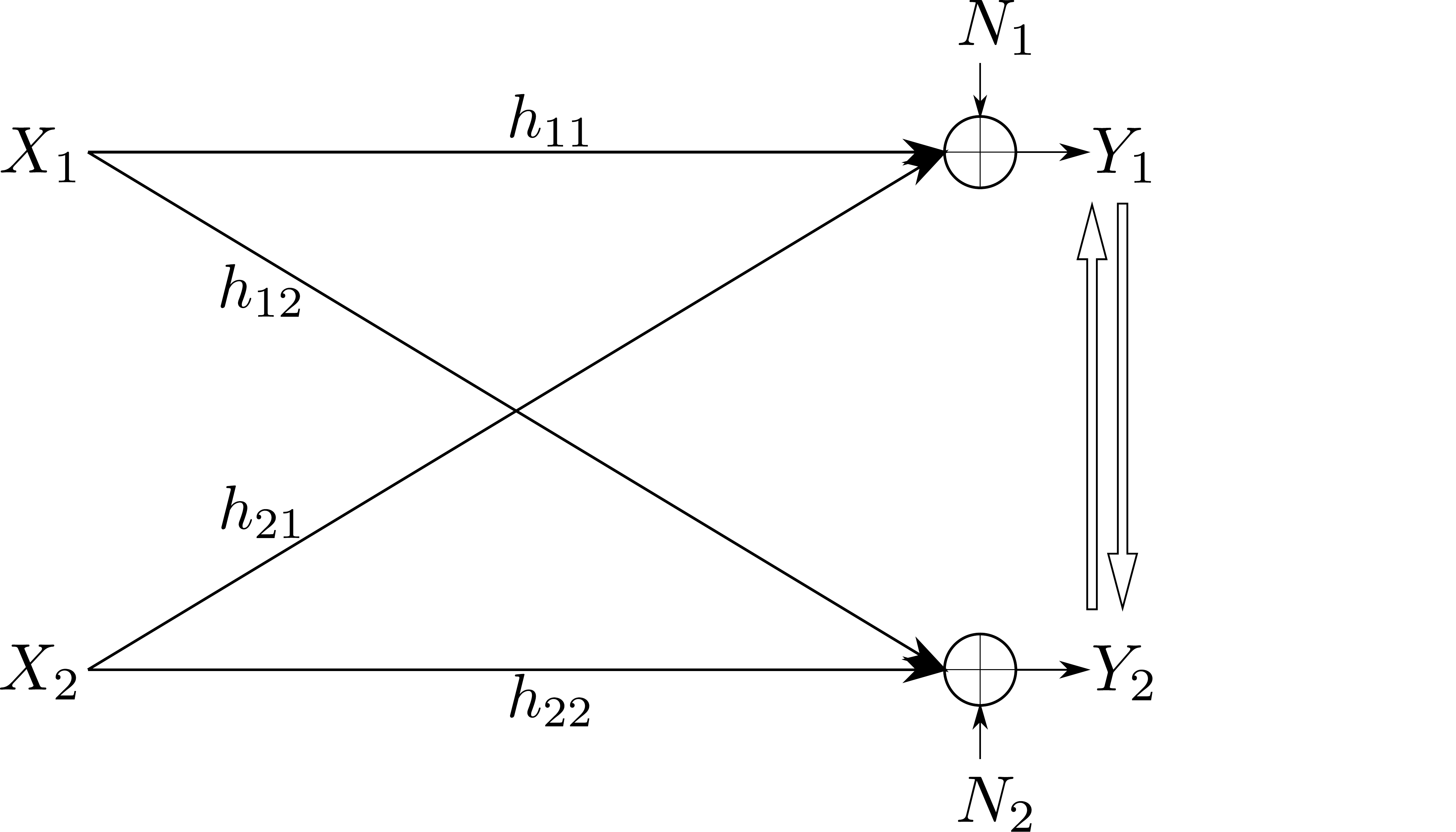}
\caption{Each user can play the role of a relay for another user. The same chessboard quantization strategy can be used for limited receiver cooperation to resolve interference. \label{fig:rcoop}}
\end{figure}

The proposed relay strategy can also be used to allow receiver cooperation in an interference channel. This is illustrated in Fig.~\ref{fig:rcoop} where each receiver is connected via a digital link of limited rate to the other. Each receiver plays the role of a relay for the other user, quantizes its observation prior to decoding using the same chessboard-like quantization strategy, and forwards the index corresponding to its quantized observation to the other user. Decoding is performed again by computing enhanced LLR values and iterative decoding.

\begin{figure}
\centering
\includegraphics[width=9cm]{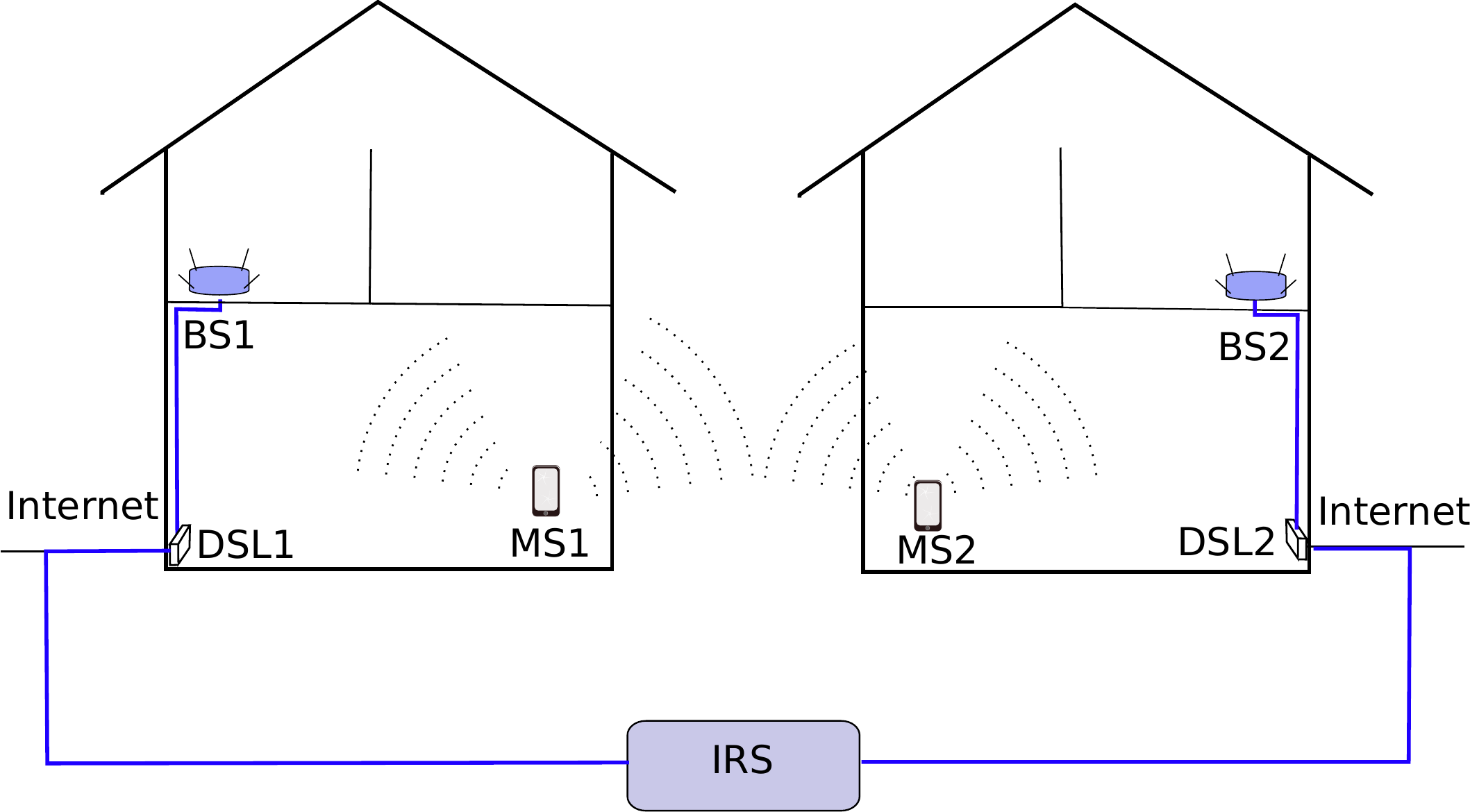}
\caption{An uplink scenario where the two users cooperate as relays for each other to reduce interference via their Internet connection. Since the relay operation in coarse network coding strategy is independent of the destination being served, and since the same relay message can be used by different destinations, an interference resolution server (IRS) can consolidate relay messages from participating base station relays. An interfered receiver queries the IRS server to receive relevant relayed information about neighboring interference sources to assist its decoder. \label{fig:rcoopbs}}
\end{figure}

Note that the analysis for the shared relay scenario shows that the same relay message can be used by different receivers; thus, generating relay messages are independent of the end receiver and the relay messages could be consolidated in a server to be used by any relevant interfered receiver. As an application, Fig.~\ref{fig:rcoopbs} shows a possible uplink scenario in a residential cellular wireless communication network where the two base stations use their Internet connections to communicate with each other as limited-rate relay nodes via a server. In the uplink, the base station playing the role of a relay shares its channel state information and bin indices for its quantized observation with neighboring base stations using, for example, an  {\em interference resolution} server. A base station experiencing interference queries the server for relayed information from neighboring interfering base stations. The receiver then incorporates the retrieved information to enhance the initial LLR calculations, which are then used by the iterative decoder. 

\appendix
In this appendix, we prove the asymptotic incremental optimality of scalar quantization/binning for a Gaussian interference channel for $R_0=1$, and with Gaussian inputs, where interference is treated as noise. That is, we prove that one bit relayed improves the rate of each user by one bit asymptotically at high SNRs.

The incremental rate improvement for user one can be found as follows:
\begin{align}
I(X_1;Y_1,X_r)&=I(X_1;Y_1)+I(X_1;X_r|Y_1)\non\\
&=I(X_1;Y_1)+I(X_1;X_r|h_{11}X_1+h_{21}X_2)\non\\
&=I(X_1;Y_1)+H(X_r|Y_1)-H(X_r|X_1,Y_1)\non\\
&=I(X_1;Y_1)+H(X_r|h_{11}X_1+h_{21}X_2+N_1)\non\\
&\hspace{1cm}-H(X_r|h_{11}X_1+h_{21}X_2+N_1,X_1)\non.
\end{align}
We prove that as $N_0\ra 0$,  
\begin{align}
H(X_r|h_{11}X_1+h_{21}X_2+N_1)\ra 1 \label{eq:y1}
\intertext{and}
H(X_r|h_{11}X_1+h_{21}X_2+N_1,X_1)\ra 0.\label{eq:x1y1}
\end{align}
 Since $X_r=\left\lfloor {Y_r}/{d}\right\rfloor \mod 2$, we first need to find the conditional distributions of $Y_r$ given $Y_1$ and $(Y_1,X_1)$. We have:
\begin{align}
p(y_r|x_1,y_1)&=\frac{1}{\pi\sigma}\exp\left\{-\frac{\norm{y_r-\mu_1}^2}{\sigma_1^2}\right\}dy_r,
\end{align}
where
\begin{align}
\sigma_1^2&=N_0+\frac{\norm{g_2}^2P_2N_0}{\norm{h_{21}}^2P_2+N_0}\non\\
\mu_1&=g_{1}x_1+\frac{g_2h_{21}^*P_2}{\norm{h_{21}}^2P_2+N_0}(y_1-h_{11}x_1)\non
\end{align}
and
\begin{align}
p(y_r|y_1)&=\frac{1}{\pi\sigma_2}\exp\left\{-\frac{\norm{y_r-\mu_2}^2}{\sigma_2^2}\right\}dy_r,
\end{align}
where
\begin{align}
\sigma_2^2&=N_0+\frac{\norm{g_1h_{21}-g_2h_{11}}^2P_1P_2+N_0(\norm{g_1}^2P_1+\norm{g_2}^2P_2)}{\norm{h_{11}}^2P_1+\norm{h_{21}}^2P_2+N_0}\non\\
\mu_2&=\frac{g_1h_{11}^*P_1+g_2h_{21}^*P_2}{\norm{h_{11}}^2P_1+\norm{h_{21}}^2P_2+N_0}y_1.\non
\end{align}
Let \[d=N_0^{\alpha},\] with $0<\alpha<0.5$. Then, we have:
\begin{align}
\var\bigl(Y_r/d\big\vert X_1,Y_1\bigr)&=N_0^{1-2\alpha}+\frac{\norm{g_2}^2P_2}{\norm{h_{21}}^2P_2+N_0}\cdot N_0^{1-2\alpha}\non\\
&\ra 0,
\end{align}
as $N_0\ra 0$ for $\alpha<1/2$. Thus, given $X_1$ and $Y_1$, $Y_r/d$ is asymptotically deterministic, and consequently, $\lfloor Y_r/d \rfloor$ and also $\lfloor Y_r/d \rfloor\mod 2$ are deterministic. This proves \eqref{eq:x1y1}.

To prove \eqref{eq:y1}, notice that:
\begin{align}
\var\bigl(Y_r/d\big\vert X_1,Y_1\bigr)&=\left(1+\frac{\norm{g_1}^2P_1+\norm{g_2}^2P_2}{\norm{h_{11}}^2P_1+\norm{h_{21}}^2P_2+N_0}\right)\cdot N_0^{1-2\alpha}\non\\
&\hspace{1cm}+\frac{\norm{g_1h_{21}-g_2h_{11}}^2P_1P_2}{\norm{h_{11}}^2P_1+\norm{h_{21}}^2P_2+N_0}\cdot N_0^{-2\alpha} \non\\
&\ra \infty,
\end{align}
as $N_0\ra 0$ for $\alpha>0$, provided that $g_1h_{21}-g_2h_{11}\neq 0$. Thus, given $Y_1$, the variance of  $Y_r/d$ asymptotically is unbounded, and hence, $\lfloor Y_r/ d \rfloor \mod 2$ tends to a Bernoulli 1/2 random variable. This proves \eqref{eq:y1}. A similar analysis proves that the rate of the second user also improves by one bit.

\bibliographystyle{IEEE}
\bibliography{IEEEabrv,reference}

\begin{thebibliography}{10}

\bibitem{zamir_shamai_erez}
R.~Zamir, S.~Shamai, and U.~Erez,
\newblock ``Nested linear/lattice codes for structured multiterminal binning,''
\newblock {\em {IEEE} Trans. Inform. Theory}, vol. 48, no. 6, pp. 1250--1276,
  June 2002.

\bibitem{wireless_network_coding}
S.\ Katti, H.\ Rahul, W.\ Hu, D.\ Katabi, M.~M\'{e}dard, and J.~Crowcroft,
\newblock ``{XOR}s in the air: Practical wireless network coding,''
\newblock {\em {IEEE/ACM} Trans. Networking}, vol. 16, no. 3, pp. 497--510,
  June 2008.

\bibitem{li_yeung_cai}
S.~Y.~R. Li, R.~W. Yeung, and N.~Cai,
\newblock ``Linear network coding,''
\newblock {\em IEEE Trans. Inform. Theory}, vol. 49, no. 2, pp. 371--381, Feb.
  2003.

\bibitem{cover_elgamal}
T.~M. Cover and A.~El Gamal,
\newblock ``Capacity theorems for the relay channel,''
\newblock {\em {IEEE} Trans. Inform. Theory}, vol. 25, no. 5, pp. 572--584,
  Sept. 1979.

\bibitem{lim_kim_elgamal}
S.~H. Lim, Y.-H. Kim, A.~El Gamal, and S.-Y. Chung,
\newblock ``Noisy network coding,'' submitted to IEEE Trans. Inform. Theory,
  march 2010,
\newblock arXiv:1002.3188.

\bibitem{peyman_yu_ita10}
P.\ Razaghi and W.\ Yu,
\newblock ``Universal relaying for the intereference channel,''
\newblock in {\em Proc.\ 2010 Inf. Theory and Applications Workshop ({ITA})},
  San Diego, CA, Feb. 2010.

\bibitem{ahlswede_network_coding}
R.~Ahlswede, N.~Cai, S.-Y.~R. Li, and R.~W.-H. Yeung,
\newblock ``Network information flow,''
\newblock {\em {IEEE} Trans. Inform. Theory}, vol. 46, no. 4, pp. 1204--1216,
  July 2000.

\bibitem{sahin_erkip_icancellation}
O.~Sahin and E.~Erkip,
\newblock ``On achievable rates for interference relay channel with
  interference cancellation,''
\newblock in {\em Proc. 41st Annual Asilomar Conference on Signals, Systems,
  and Computers}, Nov. 2007.

\bibitem{sahin_simeone_erkip}
O.~Sahin and O.~Simeone~E. Erkip,
\newblock ``Interference channel aided by an infrastructure relay,''
\newblock in {\em Proc. {IEEE} Int. Symp. on Inf. Theory ({ISIT})}, Seoul, June
  2009, pp. 2023--2027.

\bibitem{sahin_erkip}
O.~Sahin and E.~Erkip,
\newblock ``Achievable rates for the gaussian interference relay channel,''
\newblock in {\em Proc. {IEEE} Global Telecommun. Conf. ({GLOBECOM})},
  Washington, Nov. 2007, pp. 1--6.

\bibitem{maric_dabora_goldsmith}
I.~Maric, R.~Dabora, and A.~Goldsmith,
\newblock ``On the capacity of the interference channel with a relay,''
\newblock in {\em Proc. {IEEE} Int. Symp. on Inf. Theory ({ISIT})}, Toronto,
  ON, July 2008, pp. 554--558.

\bibitem{analog_network_coding}
A.~A. Zaidi, M.~N. Khormuji, S.~Yao, and M.~Skoglund,
\newblock ``Optimized analog network coding strategies for the white gaussian
  multiple-access relay channel,''
\newblock in {\em Proc. IEEE Inform. Theory Workshop (ITW)}, Taormina, Oct.
  2009, pp. 460--464.

\bibitem{compute_and_forward}
B.\ Nazer and M.\ Gastpar,
\newblock ``Compute-and-forward: Harnessing interference through structured
  codes,'' Aug. 2009,
\newblock arXiv:0908.2119v2.

\bibitem{cover_kim_deterministic}
T.~M. Cover and Y.-H. Kim,
\newblock ``Capacity of a class of deterministic relay channels,''
\newblock in {\em Proc. {IEEE} Inter. Symp. Inform. Theory ({ISIT})}, June
  2007, pp. 591--595.

\bibitem{gmi}
N.~Merhav, G.~Kaplan, A.~Lapidoth, and S.~Shamai,
\newblock ``On information rates for mismatched decoders,''
\newblock {\em {IEEE} Trans. Inform. Theory}, vol. 40, no. 6, pp. 1953--1967,
  Nov. 1994.

\bibitem{gallager_ldpc}
R.~G. Gallager,
\newblock {\em Low density parity check codes},
\newblock MIT Press, 1963.

\bibitem{ldpc-book}
T.~Richardson and R.~Urbanke,
\newblock {\em Modern Coding Theory},
\newblock Cambridge University Press, 2008.

\bibitem{shokrollahi_urbanke}
T.~Richardson, A.~Shokrollahi, and R.~L. Urbanke,
\newblock ``Design of capacity-approaching irregular low-density parity-check
  codes,''
\newblock {\em {IEEE} Trans. Inform. Theory}, vol. 47, no. 2, pp. 619--637,
  Feb. 2001.

\bibitem{chung_forney_001db}
S.-Y. Chung, G.~D. Forney, T.~J. Richardson, and R.~Urbanke,
\newblock ``On the design of low-density parity-check codes within 0.0045 {dB}
  of the {Shannon} limit,''
\newblock {\em {IEEE} Commun. Lett.}, vol. 5, no. 2, pp. 58--60, Feb. 2001.

\end{thebibliography}

\end{document}